\documentclass[twocolumn,twocolappendix]{aastex631}

\newcommand{\msun}{M_\odot}
\newcommand{\rsun}{R_\odot}

\usepackage{CJK}
\usepackage{acronym}
\usepackage{xcolor}
\usepackage{ulem}
\usepackage{bm}
\usepackage{amsmath}

\acrodef{sn}[SN]{supernova}
\acrodefplural{sn}[SNe]{supernovae}
\acrodef{ns}[NS]{neutron star}
\acrodef{lte}[LTE]{local thermodynamic equilibrium}

\begin{document}
\begin{CJK*}{UTF8}{ipxm}
\title{Supernova-induced binary-interaction-powered supernovae: a model for SN2022jli}

\correspondingauthor{Ryosuke Hirai}
\email{ryosuke.hirai@riken.jp, ryosuke.hirai@monash.edu}

\author[0000-0002-8032-8174]{Ryosuke Hirai (平井遼介)}
\affiliation{Astrophysical Big Bang Laboratory (ABBL), RIKEN Pioneering Research Institute (PRI), 2-1 Hirosawa, Wako, Saitama 351-0198, Japan}
\affiliation{School of Physics and Astronomy, Monash University, Clayton, Victoria 3800, Australia}
\affiliation{OzGrav: The ARC Centre of Excellence for Gravitational Wave Discovery, Australia}

\author[0000-0002-8338-9677]{Philipp Podsiadlowski}
\affiliation{University of Oxford, St Edmund Hall, Oxford OX1 4AR, UK}

\author[0000-0002-4338-6586]{Peter Hoeflich}
\affiliation{Department of Physics, Florida State University, Tallahassee, 32306, USA}

\author[0000-0002-0960-5407]{Maxim V. Barkov}
\affiliation{Institute of Astronomy, Russian Academy of Sciences, Moscow, 119017 Russia}

\author[0000-0002-7316-9240]{Conrad Chan}
\affiliation{Astronomy Data and Computing Services (ADACS), The Centre for Astrophysics \& Supercomputing, Swinburne University of Technology, P.O. Box 218, Hawthorn, VIC 3122, Australia}

\author[0000-0002-4856-1750]{David Liptai}
\affiliation{Astronomy Data and Computing Services (ADACS), The Centre for Astrophysics \& Supercomputing, Swinburne University of Technology, P.O. Box 218, Hawthorn, VIC 3122, Australia}

\author[0000-0002-7025-284X]{Shigehiro Nagataki (長瀧重博)}
\affiliation{Astrophysical Big Bang Laboratory (ABBL), RIKEN Pioneering Research Institute (PRI), 2-1 Hirosawa, Wako, Saitama 351-0198, Japan}
\affiliation{RIKEN Center for Interdisciplinary Theoretical and Mathematical Sciences (iTHEMS), 2-1 Hirosawa, Wako, Saitama 351-0198, Japan}
\affiliation{RIKEN-Berkeley Center, RIKEN iTHEMS, University of California, Berkeley, Berkeley, CA 94720, USA}
\affiliation{Astrophysical Big Bang Group (ABBG), Okinawa Institute of Science and Technology Graduate University (OIST), 1919-1 Tancha, Onna-son, Kunigami-gun, Okinawa 904-0495, Japan}

\begin{abstract}
We present 3D hydrodynamical modelling of supernova-induced binary-interaction-powered supernovae; a scenario proposed for the peculiar type Ic supernova SN2022jli. In this scenario, supernova ejecta of a stripped-envelope star impact a close-by stellar companion, temporarily inflating the envelope. The expanded envelope engulfs the neutron star, causing strong mass accretion at super-Eddington rates. Feedback from the accretion powers the supernova light curve with periodic undulations. Our simulations capture key features of SN2022jli, both the overall decline and the superimposed undulations of the light curve. Based on our parameter study, we find that (i) the accretion feedback should be sufficiently geometrically confined and (ii) the eccentricity of the post-supernova binary orbit should be $0.8\lesssim e\lesssim0.9$ to sustain a high accretion rate and match the low undulation amplitude ($\Delta L/L\sim0.1$) of SN2022jli. Different combinations of parameters could account for other supernovae like SN2022mop, SN2009ip and SN2015ap, which have varying undulation periods and amplitudes. We also discuss possible explanations for other key features of SN2022jli such as the $\gamma$-ray detection at $\sim200~\mathrm{d}$ and the rapid optical drop at $\sim250~\mathrm{d}$. Finally, we speculate on the future evolution of the system and its relation to existing neutron star binaries.
\end{abstract}


\section{Introduction} \label{sec:intro}

SN2022jli is a type Ic \ac{sn} that was discovered in a nearby galaxy NGC 157 on 5 May 2022 \citep[]{moo23,che24}. Following a rather normal-looking type Ic \ac{sn} for $\sim25~\mathrm{d}$, the light curve turned back up and peaked again at $\sim50~\mathrm{d}$ with a luminosity similar to that of the first peak. After this second peak, the light curve declined at a rate shallower than expected from $^{56}\mathrm{Ni}$ decay with an iconic $\sim12.5~\mathrm{d}$ periodic undulation. GeV $\gamma$-rays were detected at $\sim200~\mathrm{d}$, while no X-rays were detected around the same time \citep[]{che24}. Soon after the $\gamma$-ray detection, the optical light curve started rapidly dropping at $\sim250~\mathrm{d}$ accompanied with evidence of dust formation \citep[]{car24}. Additionally, H$\alpha$ emission was detected at late times, with possible evidence of periodically shifting radial velocity \citep[]{che24}.

The colour in the second portion of the light curve was much bluer than other known stripped-envelope \acp{sn}, indicating an extra powering mechanism \citep[]{car24}. Some studies propose magnetar spin-down as the source of the extra power \citep[]{car24,ore25}, although it requires an additional mechanism to explain the undulations as it does not naturally have any periodicity. Due to the clear periodicity, it was speculated that this \ac{sn} is binary-interaction-powered as proposed in \citet{RH22b}. In this scenario, the nascent \ac{ns} directly penetrates the outer layers of a main sequence companion star, causing strong mass accretion that powers the light curve from inside the ejecta. However, such direct interactions can only occur several times as the orbit will rapidly decay upon each penetration, whereas SN2022jli showed an almost fixed periodicity for at least $\gtrsim15$ cycles.

In this paper, we model a scenario we proposed in previous studies \citep{oga21,moo23,RH24b} with 3D hydrodynamic simulations. It builds on the scenario of \citet{RH22b} but with the additional effect of ejecta-companion interaction, where the outer layers of the companion star is heated by the impact of the \ac{sn} ejecta and forced to inflate \citep[e.g.][]{RH18}. With sufficient \ac{sn}-heating, the companion can inflate enough to overflow its Roche lobe or engulf the entire post-\ac{sn} orbit, inducing binary interactions. The density of the inflated layers are typically low enough to avoid any measurable amount of orbital decay, but simultaneously high enough to enable strong mass accretion onto the remnant compact object \citep{hob22,sok22}. Since the binary orbit is almost guaranteed to be eccentric due to the \ac{sn} mass loss and natal kick, we expect there to be periodic modulations in the accretion rate as the accretor travels through various depths of the companion envelope. The envelope is sufficiently optically thick to reprocess any radiative and kinetic feedback from the accretor into optical wavelengths. For all of the above reasons, this is an ideal situation to create bright undulating light curves like that of SN2022jli.

This paper is organized as follows. We first outline our model parameters and the methods for the hydrodynamic simulations and light curve modelling in Section~\ref{sec:method}. Simulation results are presented in Section~\ref{sec:results}, along with a comparison of the simulated light curve with the observed light curve of SN2022jli. We then discuss how to explain other key features of SN2022jli within our scenario as well as speculations on the future evolution of the system in Section~\ref{sec:discussion}. We finally summarize our results in Section~\ref{sec:summary}.

\section{Method}\label{sec:method}

\subsection{Binary parameters}

To test our model, we carry out 3D hydrodynamic simulations of the interaction between \ac{sn}-heated companions and new-born \acp{ns}\footnote{Throughout this paper we assume that the compact object is a \ac{ns}, which seems most consistent with the \ac{sn} properties and the inferred kick properties. There are also more mechanisms for \acp{ns} that allow super-Eddington accretion. However, we cannot completely rule out the possibility that it is a black hole.}. We consider a post-\ac{sn} binary that consists of a $M_\mathrm{NS}=1.4~\msun$ \ac{ns} and a stellar companion on an eccentric orbit with an orbital period of $P_\mathrm{orb}=12.5~\mathrm{d}$. The period is chosen to match the undulation period of SN2022jli \citep[e.g.][]{moo23}. We treat eccentricity as a free parameter, given the lack of observational constraints\footnote{\citet{che24} report a tentative detection of a periodic velocity shift in the late-time H$\alpha$ emission. They estimate an eccentricity of $e=0.7$--$0.96$ based on radial velocity fits. See Section~\ref{sec:eccentricity} for alternative constraints.}. The model implies that the pre-\ac{sn} orbital period was shorter, but the orbit widened and the eccentricity increased due to the combination of mass ejection and the natal kick imparted on the \ac{ns}.

There are several possible constraints on the companion mass. Since the \ac{sn} was classified as type Ic and the SN2022jli progenitor system likely had a pre-\ac{sn} orbital period of $P_\mathrm{orb}<12.5~\mathrm{d}$, it most likely experienced a common-envelope phase that stripped the primary envelope and tightened the orbit. Although the exact conditions for initiating common-envelope phases are not known \citep[e.g.][]{iva13,pav17,tem23,wil23}, it is considered that low mass ratio systems are generally more unstable. This places a loose upper bound on the companion mass to be sufficiently smaller than the pre-common-envelope primary mass. On the other hand, the companion mass should be large enough such that the binary can stay bound after the \ac{sn}. In the absence of \ac{ns} natal kicks, the binary will survive when the ejecta mass is less than half of the total mass of the binary. The estimated ejecta mass for SN2022jli is $M_\mathrm{ej}\sim1.5~\msun$ \citep[]{che24,car24}, so assuming a \ac{ns} mass of $M_\mathrm{NS}=1.4~\msun$, we cannot place a useful lower limit on the companion mass. Given these considerations, we choose a companion mass of $M_2=5~\msun$ as our fiducial model. For comparison, we also run models with companion masses of $M_2=3~\msun$ and $10~\msun$. All stellar models are constructed with the stellar evolution code MESA \citep[v24.08.1;][]{MESA1} with mostly default settings, assuming that the companion is close to its zero-age main-sequence structure.

For the orbital eccentricity, we choose $e=0.5$ as our baseline model and run $e=0.3, 0.7$ and $0.8$ for comparison. All model parameters are summarized in Table~\ref{tab:parameters} along with the periastron distance $a_\mathrm{per}$ and assumptions for the accretion and feedback (see Section~\ref{sec:feedback}). If we assume the pre-\ac{sn} orbit was circular and that the ejecta mass was $M_\mathrm{ej}=1.5~\msun$, the mass loss alone can only excite an eccentricity of $e=M_\mathrm{ej}/(M_\mathrm{NS}+M_2)\sim0.34, 0.23, 0.13$ for $M_2=3, 5, 10~\msun$ respectively. Our choices of eccentricity implies that there was also a natal kick imparted to the \ac{ns} with a magnitude of the order of $v_\mathrm{kick}\gtrsim50~\mathrm{km~s}^{-1}$.

\begin{table}
 \begin{center}
  \caption{Model parameters.\label{tab:parameters}}
  \begin{tabular}{cccccc}
   \hline
   Name & $M_2$ & $e$ & $a_\mathrm{per}$ & $p$ & Feedback\\
   & ($\msun$) & & ($\rsun$) & & \\
   \hline
   M05e05b &  5 & 0.5 & 21.1 & $0.0$ & Bipolar \\
   M05e03b &  5 & 0.3 & 29.6 & $0.0$ & Bipolar \\
   M05e07b &  5 & 0.7 & 12.7 & $0.0$ & Bipolar \\
   M10e05b & 10 & 0.5 & 25.6 & $0.0$ & Bipolar \\
   M03e05b &  3 & 0.5 & 18.7 & $0.0$ & Bipolar \\
   M05e05n &  5 & 0.5 & 21.1 & $0.0$ & None \\
   M05e05t &  5 & 0.5 & 21.1 & $0.0$ & Thermal \\
   M05e07b\_r & 5 & 0.7 & 12.7 & $0.3$ & Bipolar \\
   M05e08b\_r & 5 & 0.8 & 8.44 & $0.3$ & Bipolar \\
   \hline
  \end{tabular}
 \end{center}
\end{table}

\subsection{Hydrodynamics code}

For all simulations we use the 3D (magneto-) hydrodynamics code HORMONE \citep[]{RH16}, which is a grid-based code that solves the Euler equations based on a Godunov-type scheme. Self gravity is implemented via the hyperbolic self-gravity method \citep[]{RH16} to achieve high computational efficiency. We use an equation of state that includes the contribution of ideal gas and radiation, assuming \ac{lte} between the gas and radiation. HORMONE has recently been fully MPI-parallelized, enabling it to scale almost linearly on multiple compute nodes.

We use a spherical coordinate system for all of our simulations. The companion star model is centered on the origin. Since our code cannot handle vacuum, we attach a dilute atmosphere outside the star, with a density profile $\rho=\rho_0(R_2/r)^2$, where $\rho_0$ is set to a value lower than the surface density of the companion. The outer boundary extends out to $r_\mathrm{out}=8\times10^{15}~\mathrm{cm}\sim115,000~\rsun$, which is $\sim3000$ times larger than the orbital semimajor axis. The radial grid spacing is chosen such that it is smallest around the surface of the star and increases in a geometric series inwards and outwards. This ensures that the pressure scale height in the star is always resolved by at least $\sim10$ grid points\footnote{Except for the very surface layers where the pressure scale height is prohibitively small.}. We cover the star with $\sim300$ radial grid points, and $N_r=800$ grid points for the entire domain. We assume equatorial symmetry, and thus only simulate the upper hemisphere where the polar grid spacing is uniform in $\cos\theta$, with $N_\theta=32$ grid points from pole to equator. This ensures that the solid angle per cell is uniform, minimizing asymmetries in numerical viscosity. It also places more resolution around the orbital plane, where most of the interesting dynamics are expected to take place. The azimuthal spacing is uniform with $N_\phi=256$ grid points. To relax the severe Courant conditions around the coordinate origin, we gradually reduce the angular resolution towards the centre, assuming full spherical symmetry ($N_\theta=N_\phi=1$) in the innermost 5 cells, $(N_\theta, N_\phi)=(2, 16)$ in the next 5 cells, $(N_\theta, N_\phi)=(4, 32)$ in the next 6 cells and so on.

In order to efficiently resolve the companion star at all times, we solve the hydrodynamics on a non-inertial frame fixed to the centre of the companion. To do this, at each time step, we first compute the acceleration at the coordinate origin and then subtract that value from every cell in the computational domain. We start all simulations from periastron, assuming the \ac{sn} happened at $t=0$ and the orbit was instantaneously transformed to its current state.

\subsection{SN-heating}

After the \ac{sn} explosion, the \ac{sn} ejecta collide with the companion star, heating the surface layers through shocks. Only a small amount of mass is ejected in this process ($\lesssim0.1~\%$) and the main effect is that the injected heat drives the surface out of thermal equilibrium, causing it to become overluminous and inflated \citep[]{RH15,RH18,oga21,che23,RH23}. Depending on the amount of energy injection and the stellar mass, the star can reach radii of a few $\times100~\rsun$. This is far beyond the orbital semi-major axis, inevitably causing direct interactions between the \ac{ns} and the inflated envelope.

In our simulations, we do not simulate the \ac{sn} ejecta--companion interaction but instead inject energy into the envelope to mimic the \ac{sn}-heating effect. According to the 2D hydrodynamic simulations in \citet{RH18}, the excess specific energy distribution takes the form
\begin{equation}
 \Delta\epsilon(m)=\frac{E_\mathrm{heat}}{m_\mathrm{h}[1+\ln{(M_2/m_\mathrm{h})]}}\times
 \begin{cases}
 1, & \text{if $m\leq m_\mathrm{h}$}.\\
 m_\mathrm{h}/m, & \text{if $m>m_\mathrm{h}$}.
 \end{cases}
\label{eq:Eexcess}
\end{equation}
where $m$ is the mass from the surface, $M_2$ is the companion mass, $E_\mathrm{heat}$ is the energy injected into the star, and $m_\mathrm{h}$ is a parameter that describes the efficiently heated mass. The injected energy can be computed as $E_\mathrm{heat}=pE_\mathrm{exp}\tilde{\Omega}$, where $p$ is the energy deposition efficiency, $E_\mathrm{exp}$ is the explosion energy  and $\tilde{\Omega}$ is the fractional solid angle subtended by the companion described as
\begin{equation}
 \tilde{\Omega}\equiv\frac{1-\sqrt{1-(R_2/a)^2}}{2}\sim \frac{R_2^2}{4a^2}.
\end{equation}
 Here, $R_2$ is the companion radius and $a$ is the orbital separation. The value of the energy deposition efficiency ranges around $p\sim0.08$--$0.12$, depending on how much the star gets compressed by the \ac{sn} ejecta (see the discussion in Section~4.3 of \citealt{RH18}). We use $p=1/12\sim0.08$ for our simulations. Similarly, $m_\mathrm{h}$ is related to the ejecta mass $M_\mathrm{ej}$ through $m_\mathrm{h}=M_\mathrm{ej}\tilde{\Omega}/2$. The exact choice of $m_h$ does not influence the excess heat distribution as long as $m_h\ll M_2$.

The heating timescale is roughly of the order of the shock crossing time of the star
\begin{equation}
 \tau_\mathrm{heat}\sim210~\mathrm{s}\left(\frac{R_2}{3\,\rsun}\right)\left(\frac{v_\mathrm{ej}}{10^4~\mathrm{km~s}^{-1}}\right)^{-1},
\end{equation}
where $v_\mathrm{ej}$ is the ejecta velocity. This is much shorter than the orbital period, so we ignore this time delay and inject excess energy into the companion at $t=0$ (first periastron).

\subsection{Feedback method}\label{sec:feedback}
Accretion onto \acp{ns} is an extremely complex phenomenon involving interaction between the accreting matter and magnetic field, formation of viscous accretion disks, launching of jets and kinetic outflows, etc. Given the computational limitations, we do not attempt to resolve these phenomena, but instead treat the newly born \ac{ns} as a point particle that only interacts with the gas through gravity. We soften the gravitational potential around the particle according to the cubic-spline kernel in \citet{pri07} to avoid singularities. The softening radius is chosen to be $R_\mathrm{soft}=1~\rsun$ or 3 times the grid resolution around the particle, whichever is larger. The latter condition is chosen only close to apastron, where the local grid resolution is coarser, but also where the density is lower and hence the accretion is less important.

Accretion feedback is injected around the \ac{ns} artificially. To assess the uncertainties of our chosen feedback form, we compare results with three types of feedback; (i) no feedback, (ii) thermal feedback and (iii) bipolar kinetic feedback. All feedback is applied in an operator split manner, where all the accretion-related terms are applied after the hydrodynamics step.

The accretion and feedback is treated in our code as follows. 
For all cells within a specified accretion radius $|\bm{r}-\bm{r}_\mathrm{NS}|\leq R_\mathrm{acc}$, we assume that the mass is captured by the \ac{ns} on the local dynamical timescale. The rate of mass and angular momentum captured by the \ac{ns} is computed as
\begin{align}
    \dot{M}_\mathrm{cap}&=\int_{|\bm{r}-\bm{r}_\mathrm{NS}|\leq R_\mathrm{acc}}\frac{\rho}{t_\mathrm{ff}}dV,\\
    \dot{J}_\mathrm{cap}&=\int_{|\bm{r}-\bm{r}_\mathrm{NS}|\leq R_\mathrm{acc}}\frac{\rho}{t_\mathrm{ff}}(\bm{r}-\bm{r}_\mathrm{NS})\times(\bm{v}-\bm{v}_\mathrm{NS})dV,
\end{align}
and the local free-fall time $t_\mathrm{ff}$ is defined as
\begin{equation}
 t_\mathrm{ff}\equiv \frac{R_\mathrm{acc}}{\sqrt{\phi_\mathrm{soft}(R_\mathrm{acc})-\phi_\mathrm{soft}(|\bm{r}-\bm{r}_\mathrm{NS}|)}},
\end{equation}
where $\phi_\mathrm{soft}(r)$ is the softened gravitational potential of the \ac{ns}.

For most of our models, we optimistically assume that all of the captured material will be accreted onto the \ac{ns} (optimistic model). However, based on magneto-hydrodynamic simulations of super-Eddington accretion, only a small fraction of the infalling material will actually accrete onto the compact object \citep[e.g.][]{bla99}. Most of the material can be ejected via powerful disk winds before reaching the \ac{ns} surface, not contributing much to the accretion luminosity. Therefore, assuming $\dot{M}_\mathrm{NS}=\dot{M}_\mathrm{cap}$ is likely too optimistic by orders of magnitude. As a more realistic choice, we run additional models where the mass accretion rate is scaled down as
\begin{equation}
 \dot{M}_\mathrm{NS}=\dot{M}_\mathrm{cap}\left(\frac{R_\mathrm{NS}}{R_\mathrm{circ}}\right)^p,
\end{equation}
where $R_\mathrm{circ}\equiv\dot{J}_\mathrm{cap}^2/(GM_\mathrm{NS}\dot{M}_\mathrm{cap}^2)$ is the circularization radius of the captured material (realistic model). Here, $G$ is the gravitational constant. For the value of $p$ we choose $p=0.3$, which is within the range of expected values \citep[]{bla99,ino23}. The optimistic model is equivalent to $p=0$.

To represent accretion, we apply a source term to the continuity equation in the form
\begin{equation}
 \frac{\partial\rho}{\partial t}+\nabla\cdot(\rho\bm{v})=-\frac{\rho}{t_\mathrm{ff}}\left(\frac{R_\mathrm{NS}}{R_\mathrm{circ}}\right)^p.\label{eq:density_source_realistic}
\end{equation}
We then update the internal energy density of each cell such that the specific entropy is unchanged after applying the density source term.
With the realistic mass accretion ($p=0.3$), we find that the accretion rate is typically more than an order of magnitude lower than that of the optimistic model ($p=0$). We choose higher orbital eccentricities in the realistic models than for the optimistic models (Table~\ref{tab:parameters}), which helps to  compensate the luminosity decrease.

We apply several different methods to represent accretion feedback. We calculate the accretion luminosity as
\begin{equation}
 L_\mathrm{acc}=\frac{GM_\mathrm{NS}\dot{M}_\mathrm{NS}}{R_\mathrm{NS}},\label{eq:Lacc_def}
\end{equation}
where we choose the \ac{ns} radius to be $R_\mathrm{NS}=12~\mathrm{km}$.

In the thermal feedback model, we assume that all of the accretion energy is liberated in the form of radiation, and a fraction of it is captured by the gas and thermalizes. We define the optical depth within the accretion radius as
\begin{equation}
 \tau_\mathrm{acc}=\kappa\bar{\rho}_\mathrm{acc}R_\mathrm{acc},
\end{equation}
where $\kappa$ is the opacity and $\bar{\rho}_\mathrm{acc}$ is the average density within the accretion radius. The energy captured by the gas is estimated as
\begin{equation}
 L_\mathrm{therm}=(1-e^{-\tau_\mathrm{acc}}) L_\mathrm{acc}.\label{eq:Lacc}
\end{equation}
This luminosity is added uniformly within the accretion radius as a source term to the energy equation. For the value of opacity, we choose $\kappa=0.1~\mathrm{cm}^2~\mathrm{g}^{-1}$.

In the bipolar feedback model, we assume that the accreted material forms an accretion disk that converts all of the accretion energy into bipolar outflows through jets and/or disk winds. At each time step, we first integrate the mass $M_\mathrm{cone}$ and total energy $E_\mathrm{cone}$ in the cells that are within the accretion radius ($|\bm{r}-\bm{r}_\mathrm{NS}|\leq R_\mathrm{acc}$) and satisfy
\begin{equation}
\cos^{-1}\left(\frac{|(\bm{r}-\bm{r}_\mathrm{NS})\cdot\hat{\bm{n}}_\mathrm{jet}|}{|\bm{r}-\bm{r}_\mathrm{NS}|}\right)\leq\theta_\mathrm{jet},
\end{equation}
where $\hat{\bm{n}}_\mathrm{jet}$ is a unit vector aligned with the kinetic outflow (here we assume it is aligned with the orbital angular momentum) and $\theta_\mathrm{jet}$ is the opening angle of the outflow. Then for the same cells, we replace the velocity with $v_\mathrm{jet}(\bm{r}-\bm{r}_\mathrm{NS})/|\bm{r}-\bm{r}_\mathrm{NS}|$, where $v_\mathrm{jet}\equiv\sqrt{2(E_\mathrm{cone}+L_\mathrm{acc}\Delta t)/M_\mathrm{cone}}$ and $\Delta t$ is the time step. For the models we present, we choose $\theta_\mathrm{jet}=45^\circ$, although we tested that the results are not sensitive to the opening angle at least within the range $30^\circ\leq\theta_\mathrm{jet}\leq45^\circ$.

\subsection{Light curve modelling}

We use the post-processing module developed for HORMONE in \citet{gri21} to compute synthetic light curves. It follows the ray-tracing method of \citet{suz16} where we solve the radiative transfer equations $dI_{\nu}/d\tau=-I_{\nu}+S_\nu$, where $I_\nu, S_\nu$ are the frequency-dependent intensity and source function respectively, along rays pointing towards the observer. We assume $S_\nu=B_\nu(T)$, where $B_\nu$ is the Planck function for the local temperature and ignore any contributions of light scattered into the line of sight. The intensity is integrated over the whole projected area of the simulation domain and multiplied by $4\pi$ to obtain the isotropic-equivalent luminosity $L$. We separately carry out the same ray-tracing procedure for specific intensities at 6 different frequencies to get a rough spectrum. We fit the spectrum with a black body to compute an effective temperature $T_\mathrm{eff}$, which is then used to define a photospheric radius $R_\mathrm{eff}=\sqrt{L/(4\pi\sigma T_\mathrm{eff}^4)}$ where $\sigma$ is the Stefan-Boltzmann constant. 

For the opacity, we use an approximate analytical formula similar to \citet{met17}. See Appendix~\ref{app:opacity} for details of our opacity formula.
A key component is the floor opacity, which is a commonly used method to crudely account for other unaccounted sources such as line opacities in velocity gradients, non-\ac{lte} effects, etc \citep[e.g.][]{her90,ber11}. To bracket the range of uncertainties, we compared two calculations with $\kappa_\mathrm{floor}=0.01$ and $0.1~\mathrm{cm^2~g}^{-1}$, but find that the results are mostly the same. For the rest of this paper, we show results for $\kappa_\mathrm{floor}=0.1~\mathrm{cm^2~g}^{-1}$.

Given the high asymmetry of our problem, we also investigate the viewing direction dependence. Defining a Cartesian coordinate system with the orbital angular momentum vector as the $z$ axis, the direction opposite of the orbital eccentricity vector (Laplace-Runge-Lenz vector) as the $x$ axis, and the direction of the initial \ac{ns} velocity as the $y$ axis, we choose 5 different viewing directions; $+z, +x, -x, +y, -y$ (see the arrows in Figure~\ref{fig:snapshots_nofeedback}). Note that the $+z$ and $-z$ directions are identical as we assume equatorial symmetry. For the $\pm x, \pm y$  directions, we compute the light curve from two different polar angles, $\theta=\pi/2$ (edge on) and $\pi/4$ (off plane).

We do not simulate the \ac{sn} explosion itself, which would be contributing significantly to the light curve, especially at early times. In reality the \ac{sn} ejecta could also partially blanket the emission from the binary interactions we simulate, obscuring and/or smearing out short timescale features. To keep our results independent of the assumptions on the uncertain \ac{sn} ejecta properties, we compute the light curves ignoring any effects of the \ac{sn} ejecta. We discuss implications of this in Section~\ref{sec:delay_discussion}.

\section{Results}\label{sec:results}

\subsection{Accretion dynamics}

\begin{figure*}
 \centering
 \includegraphics[width=0.75\linewidth]{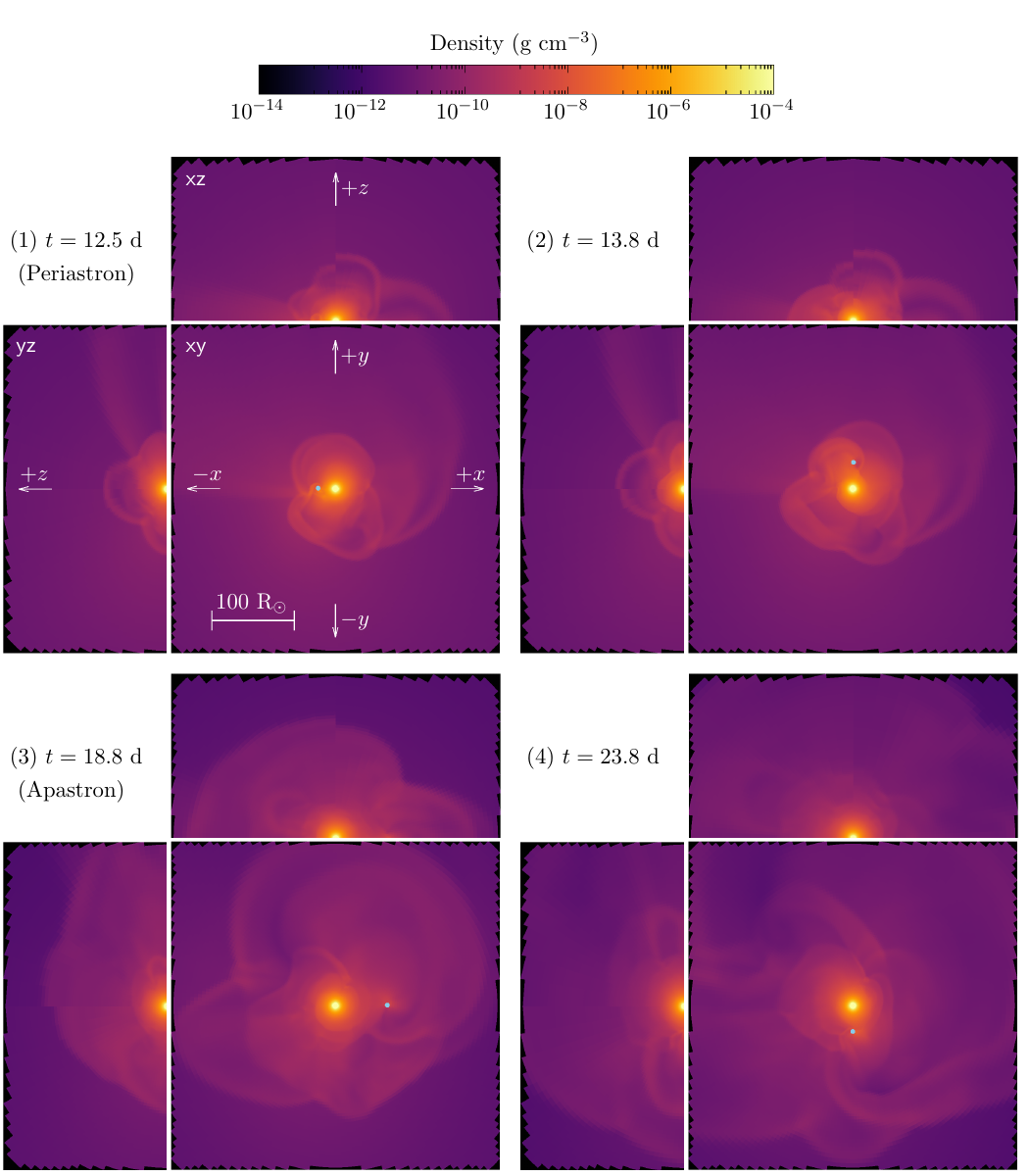} 
 \caption{Animation of the 3D hydrodynamic simulations for the no feedback model (M05e05n). We show several different slices ($xy, xz, yz$), all cut through the centre of the companion star. The $z$ axis is the direction of the orbital angular momentum and the $x$ axis is aligned with the orbital major axis. The light blue dot marks the location of the point particle representing the \ac{ns}. The animation runs from the start to end of our simulation ($\sim100~\mathrm{d}$). The static version displays four select snapshots from the second orbit. \label{fig:snapshots_nofeedback}}
\end{figure*}

\begin{figure*}
 \centering
 \includegraphics[width=0.75\linewidth]{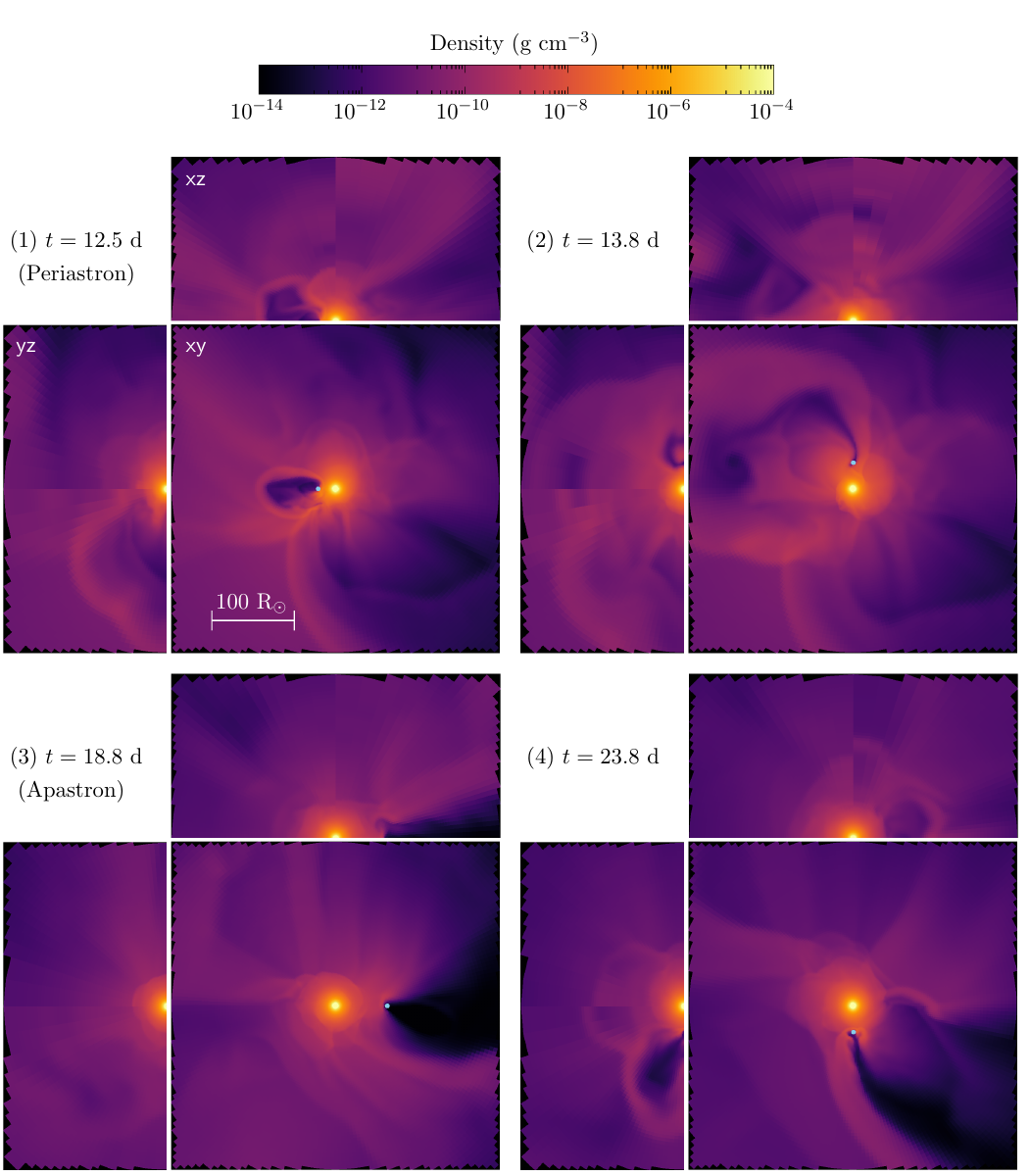} 
 \caption{Same as Figure~\ref{fig:snapshots_nofeedback} but for the thermal feedback model (M05e05t).\label{fig:snapshots_thermal}}
\end{figure*}

\begin{figure*}
 \centering
 \includegraphics[width=0.75\linewidth]{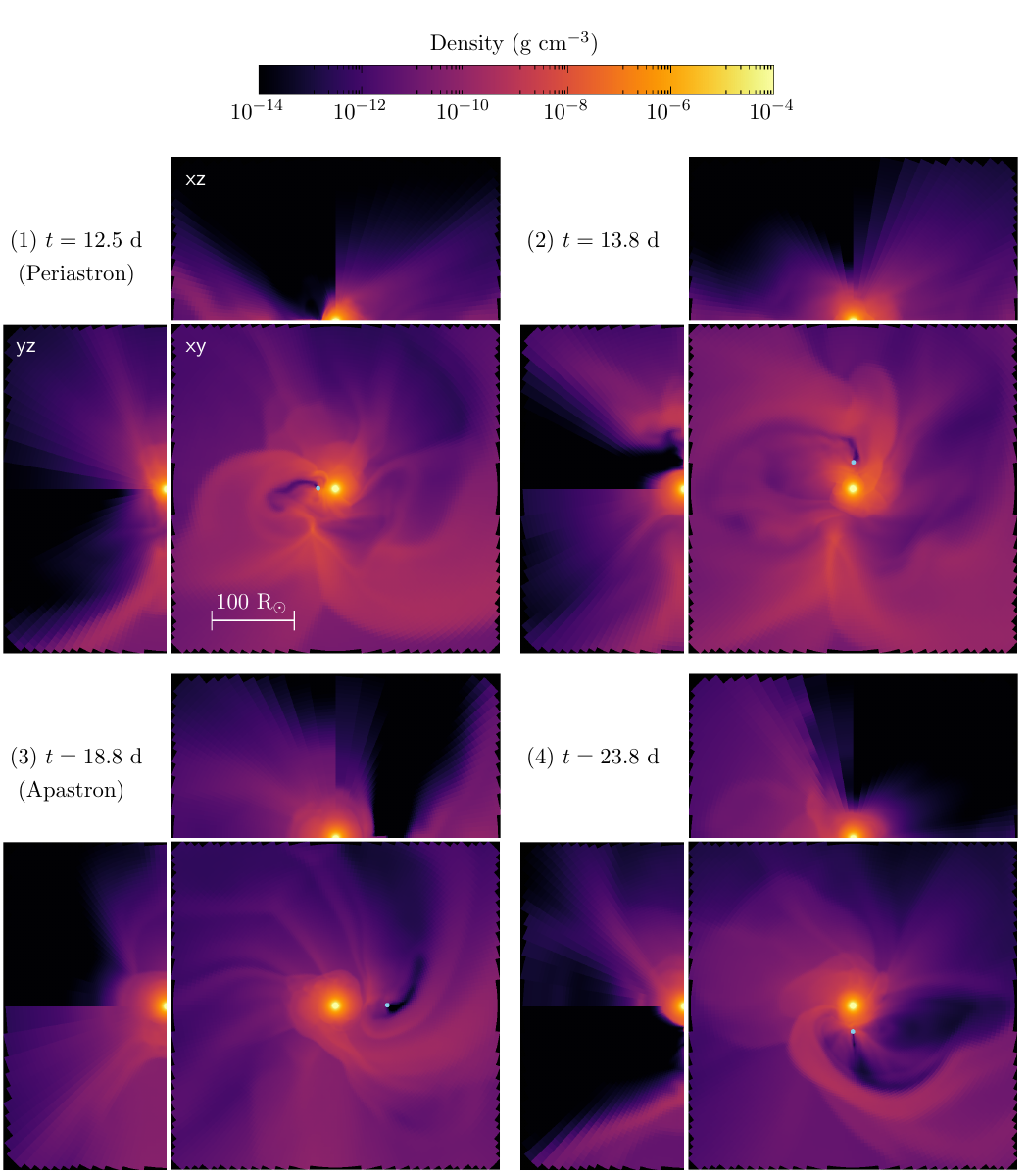}
 \caption{Same as Figure~\ref{fig:snapshots_nofeedback} but for the bipolar feedback model (M05e05b).\label{fig:snapshots_fiducial}}
\end{figure*}

\begin{figure}
 \centering
 \includegraphics[width=\linewidth]{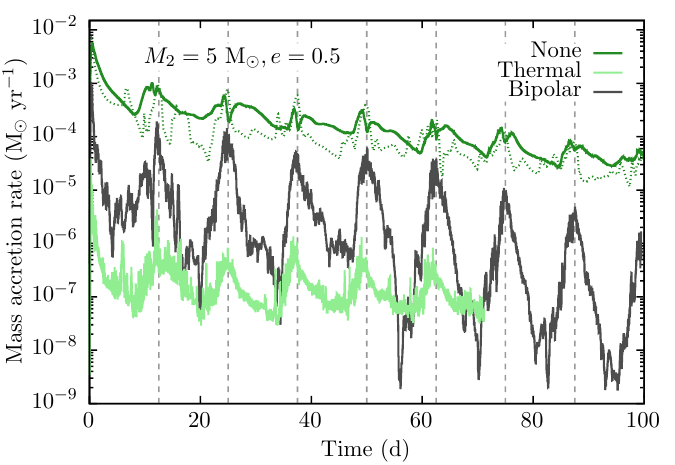}
 \caption{Mass accretion rate evolution in our hydrodynamic simulations. We compare three different models with the same orbital parameters but different feedback methods: no feedback (model M05e05n; green), thermal feedback (model M05e05t; light green) and bipolar feedback (M05e05b; dark grey). Dashed vertical lines mark the periastron passage timings. For the no feedback model, we overplot the Bondi-Hoyle-Lyttleton accretion rate computed from Eq.~(\ref{eq:bondi}) (dotted).\label{fig:mdot_feedback}}
\end{figure}

In Figures~\ref{fig:snapshots_nofeedback}--\ref{fig:snapshots_fiducial}, we display snapshots of our hydrodynamical simulations. Immediately after the start of the simulation, the companion star swells up in response to the energy injection representing \ac{sn} ejecta--companion interaction. The inflated envelope extends far beyond the location of the \ac{ns}, creating a high-density environment around it. This leads to an extremely high mass accretion rate onto the \ac{ns}, which is plotted in Figure~\ref{fig:mdot_feedback}. At the same time, there is almost no orbital decay from the dynamical friction between the envelope gas and \ac{ns} particle due to the relatively small amount of mass in the inflated envelope ($\sim10^{-4}$--$10^{-2}~\msun$).

In the no feedback model (Figure~\ref{fig:snapshots_nofeedback}), the inflated envelope is stirred up by the orbiting \ac{ns}, creating shocks in the wake. Although part of the envelope material is sucked out of the simulation through accretion, the high density around the \ac{ns} is maintained. Initially, the accretion rate onto the \ac{ns} reaches as high as $\dot{M}\sim10^{-2}~\msun~\mathrm{yr}^{-1}$, but slowly declines over time by $\sim30~\%$ per orbit. Within each orbit, as the \ac{ns} orbits through different depths of the envelope, the mass accretion rate (Figure~\ref{fig:mdot_feedback}, dark green curve) fluctuates with an amplitude of factor $\sim5$. The accretion rate sharply rises towards each periastron and slowly declines towards apastron, loosely resembling the saw-tooth shape in the SN2022jli light curve.

Overplotted in Figure~\ref{fig:mdot_feedback} is the Bondi-Hoyle-Lyttleton accretion rate computed as
\begin{equation}
 \dot{M}_\mathrm{BHL}=\frac{\pi G^2M_\mathrm{NS}^2\rho_\infty}{(v_\mathrm{NS}^2+c_{s,\infty}^2)^{3/2}},\label{eq:bondi}
\end{equation}
where $v_\mathrm{NS}$ is the \ac{ns} velocity relative to the companion star \citep[]{hoy39,bon44}. For the density and sound speed $\rho_\infty, c_{s,\infty}$, we use values on the opposite side of the star at the same radius as the \ac{ns} to obtain representative undisturbed values of the inflated envelope. Although our method does not prescribe the Bondi-Hoyle-Lyttleton rate, our accretion rates closely follow it especially around periastron.

In the thermal feedback model (Figure~\ref{fig:snapshots_thermal}), a low-density cavity is formed behind the \ac{ns}. It is roughly spherical around periastron (panel (1)), but elongates into a large cone at and after apastron (panels (3)--(4)). The cavity severely quenches the accretion rate, by $\sim4$--$5$ orders of magnitude, down to around the Eddington limit (Figure~\ref{fig:mdot_feedback}, light green curve). The effect of such ``negative feedback'' has been discussed in depth in the context of common-envelope evolution \citep[e.g.][]{sok16,gric21,hil22}.

In the bipolar feedback model (Figure~\ref{fig:snapshots_fiducial}), a low-density cavity is created in the vertical direction, while the density on the orbital plane is sustained high. The accretion rate is similarly quenched compared to the no feedback model, but only by an order of magnitude around periastron. Around apastron, the accretion is almost completely shut off similar to the thermal feedback model. As a result, the fluctuation amplitude of the mass accretion rate reaches $\sim3$--$6$ orders of magnitude. The peak accretion rates steadily decline for each periastron passage, at a rate similar to the no feedback model. 

\begin{figure}
 \centering
 \includegraphics[width=\linewidth]{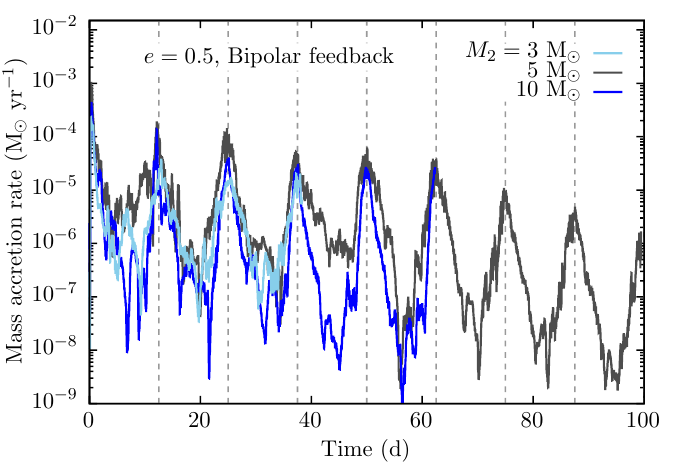}
 \caption{Same as Figure~\ref{fig:mdot_feedback} but comparing models with different companion masses: $M_2=3~\msun$ (model M03e05b; sky blue), $5~\msun$ (model M05e05b; dark grey) and $10~\msun$ (model M10e05b; blue). All models were simulated with bipolar feedback.\label{fig:mdot_mass}}
\end{figure}

Figure~\ref{fig:mdot_mass} compares the accretion rate evolution for three models with different companion star masses. All models were computed with bipolar feedback and an orbital eccentricity $e=0.5$. They all qualitatively show the same behaviour, where there are periodic fluctuations to the accretion rate with a steady decline in the long term. There seems to be no obvious trend with mass, with the $5~\msun$ model showing the highest average accretion rate compared to the $3$ and $10~\msun$ models. 

\begin{figure}
 \centering
 \includegraphics[width=\linewidth]{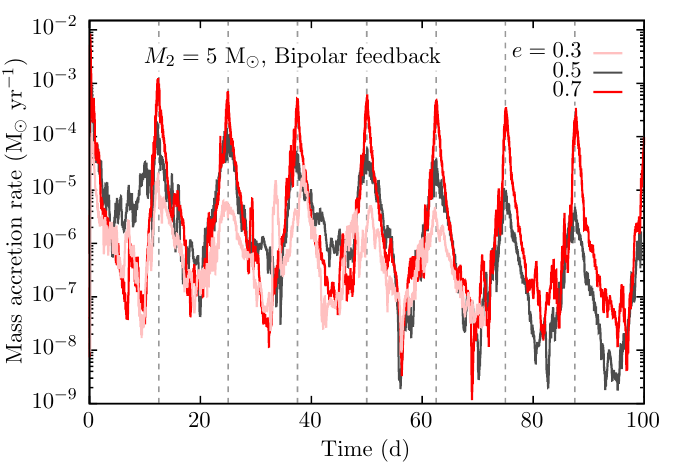}
 \caption{Same as Figure~\ref{fig:mdot_feedback} but comparing models with different eccentricities: $e=0.3$ (model M05e03b; pink), $0.5$ (model M05e05b; dark grey) and $0.7$ (model M05e07b; red). All models were simulated with bipolar feedback.\label{fig:mdot_eccentricity}}
\end{figure}

Orbital eccentricity has a stronger influence on the accretion rate. As we can see in Figure~\ref{fig:mdot_eccentricity}, the peak accretion rates at periastron clearly correlate with eccentricity, where the $e=0.7$ model has almost an order of magnitude higher accretion rate than the $e=0.5$ model. Interestingly, the $e=0.3$ model seems to lose the clear periodicity and instead shows large fluctuations that are less obviously related to the orbital phase. Note that there is a two-fold effect of varying eccentricity in our setup. Since we fix the orbital period to $P_\mathrm{orb}=12.5~\mathrm{d}$, the eccentricity determines the periastron distance and hence the pre-\ac{sn} orbital separation. Higher eccentricity leads to more energy injection through ejecta-companion interaction (for a given explosion energy), as the solid angle subtended by the companion is larger. Therefore, the higher eccentricity models lead to larger inflated radii with more mass in the inflated envelope. The other effect of higher eccentricity is that with shorter periastron distances, the \ac{ns} will orbit in deeper parts of the envelope. Even if the envelope profiles were the same, the higher eccentricity models would have higher mass accretion rates because the \ac{ns} reaches down to the deeper, higher density layers.

\begin{figure}
 \centering
 \includegraphics[width=\linewidth]{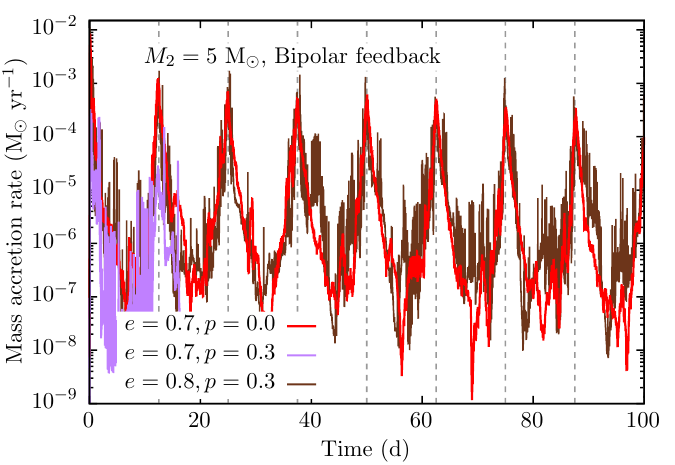}
 \caption{Same as Figure~\ref{fig:mdot_feedback} but comparing the optimistic (model M05e07b; red) vs realistic (model M05e07b\_r; purple, M05e08b\_r; dark brown) models for the mass accretion rate.\label{fig:mdot_lacc}}
\end{figure}

Figure~\ref{fig:mdot_lacc} compares the optimistic model and the realistic model for the mass accretion rate. In a short test simulation (purple curve), we found that the realistic model shows $\gtrsim15$ times lower accretion rates at periastron compared to the optimistic model with the same orbital parameters (red curve). The feedback luminosity is therefore far too weak to match that of SN2022jli. A model with higher eccentricity $e=0.8$ (black curve) roughly compensates for this loss, showing very similar mass accretion rate evolution as the $e=0.7$ model with optimistic mass accretion. Therefore, in terms of mass accretion rate evolution, there exists a degeneracy between the mass accretion efficiency ($p$) and the orbital eccentricity.

\begin{figure}
 \centering
 \includegraphics[width=\linewidth]{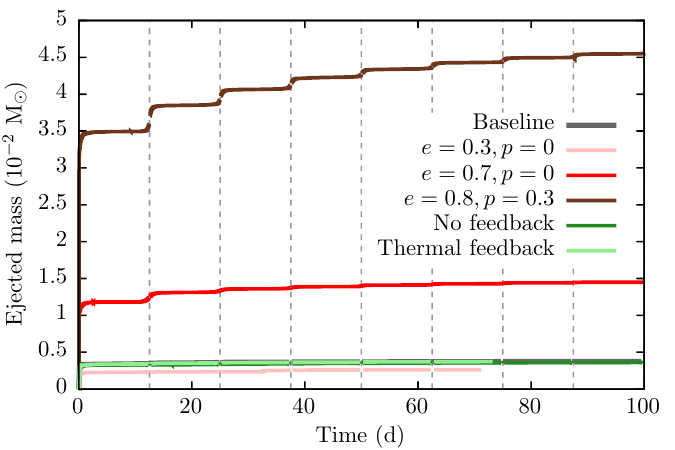}
 \caption{Time evolution of the mass ejected from the donor star. We compare all models with $M_2=5~\msun$. We treat model M05e05b as the ``Baseline'' model and colour variations are the same as in Figures~\ref{fig:mdot_feedback}, \ref{fig:mdot_eccentricity} and \ref{fig:mdot_lacc}.\label{fig:ejected_mass}}
\end{figure}

We also compare the mass ejected from the companion star in Figure~\ref{fig:ejected_mass}. To compute the companion mass, we integrate the mass in all bound cells, where a cell is marked as bound when $\phi_\mathrm{gas}+v^2/2<0$. Here, $\phi_\mathrm{gas}$ is the gravitational potential of the gas that does not include the contribution from the \ac{ns} particle, and $v\equiv|\bm{v}|$ is the cell velocity in the frame of the companion. We see that straight after the \ac{sn}, $10^{-3}$--$10^{-2}~\msun$ of mass is ejected from the companion due to the \ac{sn}-heating. While only $2\times10^{-3}~\msun$ is ejected in the $e=0.3$ model (pink curve), $1.2\times10^{-2}~\msun$ is ejected in the $e=0.7$ model (red curve) and $3.5\times10^{-2}~\msun$ in the $e=0.8$ model (dark brown curve) due to the larger energy injection via ejecta-companion interaction.
The curves with the same orbital parameters but different feedback methods (grey, dark green and light green curves) show almost identical initial mass loss (Figure~\ref{fig:mdot_feedback}). This confirms that the initial jump is purely due to \ac{sn}-heating and not because of subsequent energy injection from the feedback. After the second periastron passage, the ejected mass show small deviations between the feedback and no-feedback model, indicating that accretion feedback is causing more mass ejection from the system. Given that the total mass accretion over the duration of our simulations is $\Delta M_\mathrm{NS}\lesssim10^{-4}~\msun$, we can safely say that this mass is ejected from the system and not accreted onto the \ac{ns}. There is hardly any mass lost in the no-feedback model, indicating that tidal stripping is negligible on these timescales. In the high eccentricity models ($e=0.7, 0.8$), we can see more significant mass ejection episodes around each periastron passage. These results indicate that the depth of energy injection is important in determining the ejected mass as well as the amount.

\subsection{Light curve}

\begin{figure}
 \centering
 \includegraphics[width=\linewidth]{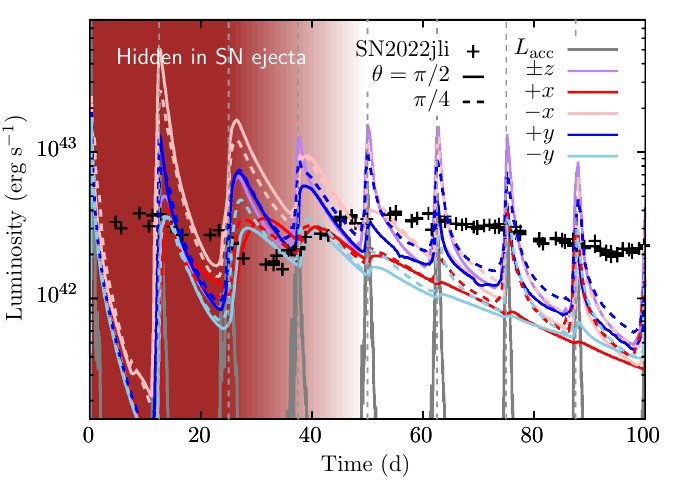}
 \caption{Simulated light curves for the $e=0.7$ model (M05e07b). Each coloured curve shows the isotropic-equivalent bolometric luminosity from various viewing directions as defined in Figure~\ref{fig:snapshots_nofeedback}. Dashed curves show the light curve from intermediate viewing angles $\theta=\pi/4$ ($45^\circ$ off the orbital plane) of the corresponding directions. The grey curve shows the accretion luminosity from the \ac{ns} particle (Eq.~(\ref{eq:Lacc})) and the black crosses are the observed light curve of SN2022jli \citep[]{che24}. The brown shaded region covers the phase where we expect the main \ac{sn} ejecta to be optically thick and thus our accretion-powered light curve is still not visible (see Section~\ref{sec:delay_discussion}).\label{fig:lightcurve_e07}}
\end{figure}

Figure~\ref{fig:lightcurve_e07} displays our synthetic light curves based on the M05e07b model, which uses the optimistic mass accretion rate and an orbital eccentricity $e=0.7$. The grey curve shows the accretion luminosity from the \ac{ns} particle, which is directly proportional to the mass accretion rate (Eq.~(\ref{eq:Lacc})). Due to the strong phase-dependence of the accretion, the energy injection into the system can be considered as an almost impulsive explosion at each periastron passage. The resulting light curves have large variations depending on the viewing direction, reflecting the complex geometry of the source binary.

Very roughly, the overall light curve shows a steady decline after the first $1$--$2$ orbits. On top of that, there are strong periodic undulations peaking at periastron. Each successive peak generally gets dimmer and dimmer for two reasons. One is that the accretion rate declines every periastron due to the decline in envelope density. Another effect is the build-up of equatorial obscuring material due to multiple mass ejections at periastron passage. Such mass ejections only occur along the orbital plane, and conversely in the poleward direction ($\pm z$), the material gets more and more evacuated by the bipolar outflows. This leads to a rise in the peak luminosities as we can see in the purple curve.

Right after the \ac{sn} explosion, there is almost no viewing direction dependence. This is because initially, the companion envelope that reprocesses the injected energy expands in an almost spherical manner due to our SN-heating method. After the second periastron passage, the light curves start to significantly deviate from each other due to the envelope being distorted by the orbital interaction. During the first few orbits, the peak luminosities are highest in the $-x$ (pink curve) and $+y$ (blue curve) directions. These angles respectively view the binary from the equatorial plane in the direction of periastron ($-x$) and the direction perpendicular to that on the post-periastron side ($+y$). From this side of the orbit, the \ac{ns} is in front of the companion star when the accretion rate (and therefore the feedback) is high. On the other side of the orbit ($+x$ and $-y$ directions), the \ac{ns} is \textit{behind} the companion when the accretion rate is high, so the radiation is damped by the optically thick inflated envelope before it reaches the observer.

All light curves show some level of undulation, where the amplitude strongly depends on the viewing direction. Generally speaking, both the peak luminosity and undulation amplitude is smaller when viewed from the orbital plane compared to the poles. When viewed from intermediate viewing angles (dashed curves), the light curve is generally in between the pole-on light curve ($\pm z$) and the equatorial light curve of the given azimuthal angle $\varphi$.

\begin{figure}
 \centering
 \includegraphics[width=\linewidth]{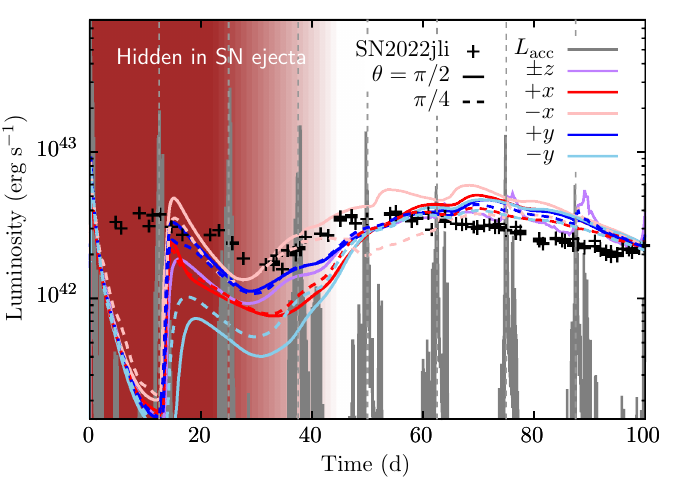}
 \caption{Same as Figure~\ref{fig:lightcurve_e08} but for the realistic mass accretion model.\label{fig:lightcurve_e08}}
\end{figure}

Figure~\ref{fig:lightcurve_e08} displays the light curves computed for Model M05e08b\_r, which uses the realistic mass accretion model and $e=0.8$. There is a striking difference to that of the M05e07b model light curve. First, the overall light curve slowly rises over a timescale of $\sim60~\mathrm{d}$ before showing a steady decline. Second, the undulation amplitudes are significantly smaller over all viewing directions. As seen in Figure~\ref{fig:mdot_lacc}, the mass accretion rate histories are very similar between these models, so the main difference is in the orbital configuration. In the higher eccentricity model, the periastron passage occurs at deeper layers inside the inflated envelope. Therefore, the feedback outflow interacts with more material as it expands, diminishing its energy before being released at the photosphere. The bipolar outflow is also smeared out over a larger solid angle, reducing the viewing direction dependency of the light curve. 

Overall, our light curves computed with the realistic mass accretion rates resemble the key features of the SN2022jli light curve. After $\gtrsim60~\mathrm{d}$, the light curve reaches a state where it steadily declines with some periodic undulations. This is likely the most important characteristic of SN2022jli that sets it apart from other known transients. Typical \acp{sn} show declines but without periodic undulations. Some mass-transferring binaries, such as X-ray binaries, show periodic modulations without a long-term decline. The combination of both the decline and undulation is naturally achieved in our model. Quantitatively, the bolometric luminosity is in the same ballpark as that of SN2022jli and so is the decline rate. The undulation amplitude depends strongly on the eccentricity and to some degree on the viewing direction, but our $e=0.8$ light curves are in good agreement with SN2022jli. On the other hand, there are some key features that we do not reproduce, including the first $\lesssim30~\mathrm{d}$ and the drop at $\sim250~\mathrm{d}$. We discuss these points in Section~\ref{sec:discussion}.

\subsection{Photospheric evolution}

\begin{figure}
 \centering
 \includegraphics[width=\linewidth]{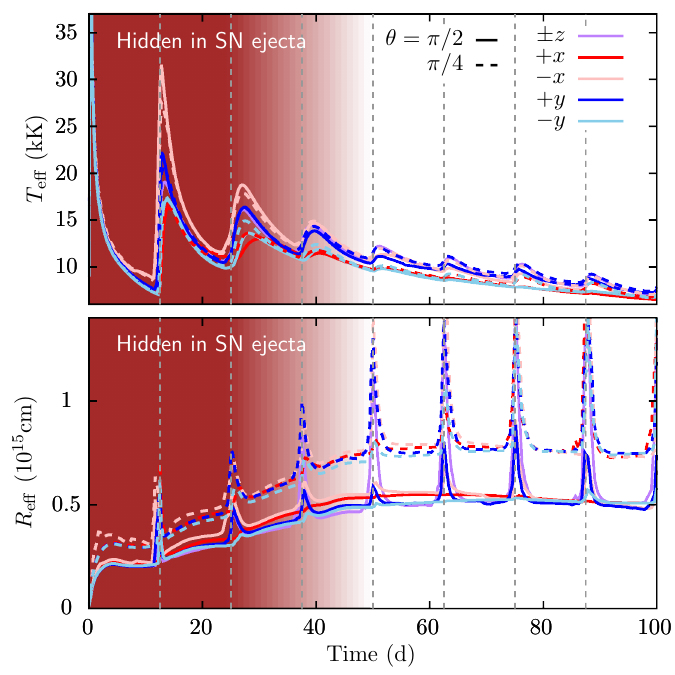}
 \caption{(Upper panel) Effective temperature from our light curve simulations for various viewing directions. (Lower panel) Photospheric radius computed from the simulations. Colours of the curves indicate the same viewing directions as in Figure~\ref{fig:lightcurve_e07}.\label{fig:Teff_Rphot_e07}}
\end{figure}

\begin{figure}
 \centering
 \includegraphics[width=\linewidth]{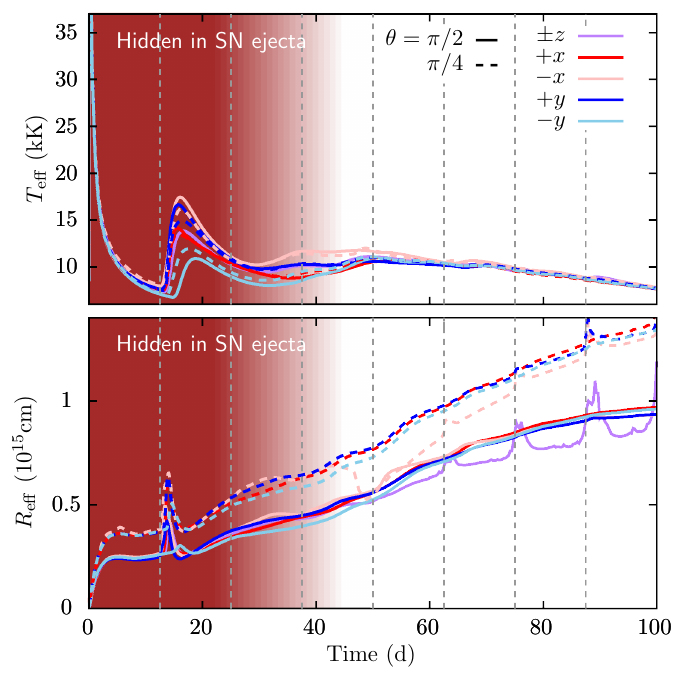}
 \caption{Same as Figure~\ref{fig:Teff_Rphot_e07} but for the realistic mass accretion model.\label{fig:Teff_Rphot_e08}}
\end{figure}

Figures~\ref{fig:Teff_Rphot_e07} \& \ref{fig:Teff_Rphot_e08} display the evolution of effective temperature and photospheric radius for models M05e07b and M05e08b\_r, respectively. In the lower eccentricity model (M05e07b), the effective temperature steadily decreases while the photospheric radius gradually increases over time in our models. There are periodic signatures both in the temperature and radius, which become weaker for the edge-on viewing directions and/or higher eccentricity. The rise in photospheric radius stalls later on, and may be starting to decline towards the end of our simulation. This plateau-like behaviour is due to the opacity having a steep drop around the recombination temperature $\sim10^4~\mathrm{K}$. Both temperature and radius roughly have the same order of magnitude as observed in SN2022jli ($T_\mathrm{eff}\sim10^4~\mathrm{K}, R_\mathrm{eff}\sim10^{15}~\mathrm{cm}$). The photospheric radius is still well contained within our computational domain ($8\times10^{15}~\mathrm{cm}$), validating our choice of the outer boundary radius. While we do not see the overall blue-ward evolution seen in SN2022jli, the model reproduces the correct phase evolution where the colour is bluer at undulation peaks \citep[]{car24}.

\begin{figure}
 \centering
 \includegraphics[width=\linewidth]{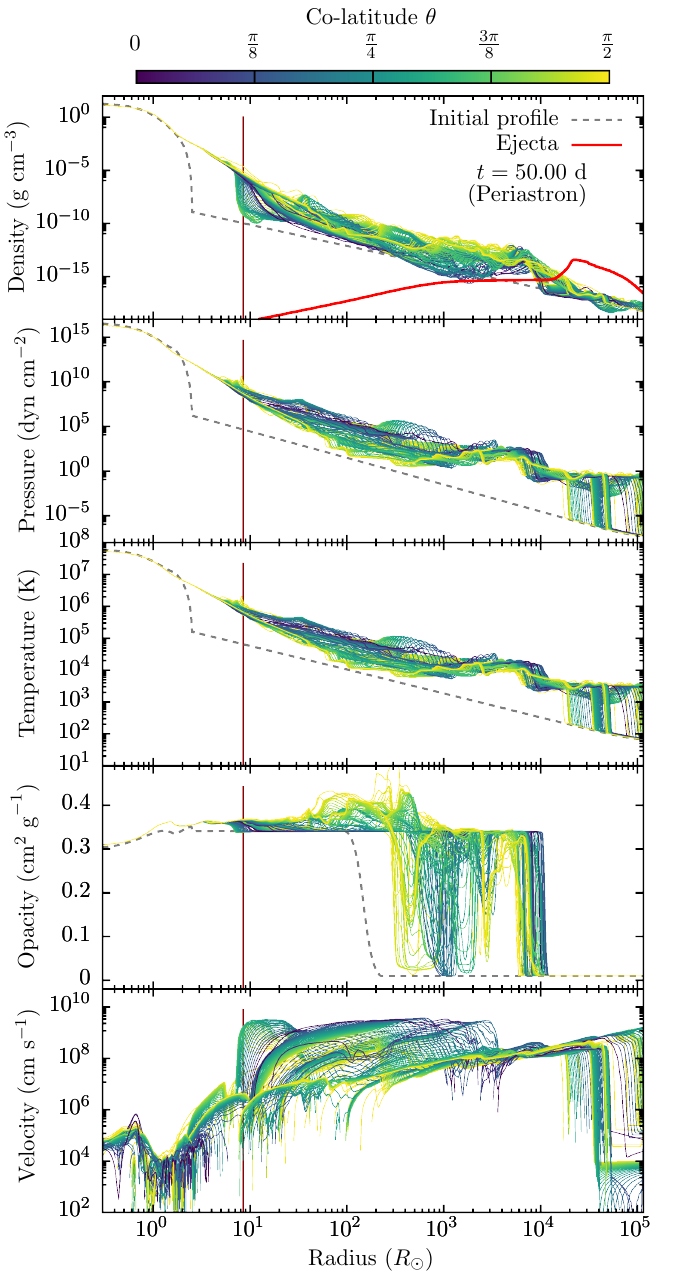}
 \caption{Profiles of density, pressure, temperature, opacity and radial velocity as a function of distance from the companion core. Each curve corresponds to the density profile along a specific radial ray and are coloured according to the latitudinal angle. The brown vertical lines mark the location of the \ac{ns}. We choose a snapshot from model M05e08b\_r at $t=50~\mathrm{d}$, which is the fourth periastron. The grey dashed curves show the initial profiles. The overlaid red solid curve is an example ejecta density profile of a type Ic \ac{sn}.\label{fig:1d_profile_periastron}}
\end{figure}

\begin{figure}
 \centering
 \includegraphics[width=\linewidth]{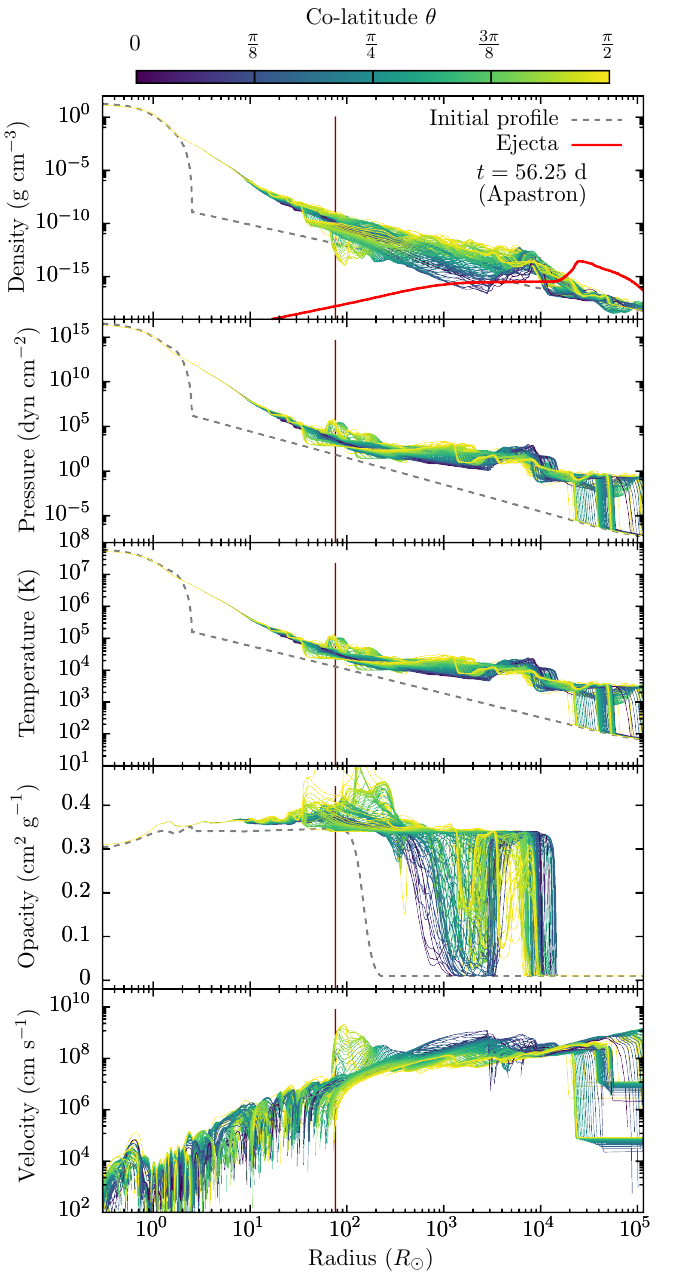}
 \caption{Same as Figure~\ref{fig:1d_profile_periastron} but for $t=56.25~\mathrm{d}$; the fifth apastron.\label{fig:1d_profile_apastron}}
\end{figure}

The complex structure of the photosphere can be viewed more clearly in Figures~\ref{fig:1d_profile_periastron} \& \ref{fig:1d_profile_apastron}. Here, we plot 1D profiles along various radial rays at given snapshots. The curves are colour-coded based on the latitudinal angle, with darker colours towards the pole and lighter colours towards the equator. We can clearly see the surface layers of the star have been significantly inflated from the initial profile (grey dashed curve), extending out to the location of the \ac{ns} and beyond.

At periastron (Figure~\ref{fig:1d_profile_periastron}), we can see the low-$\theta$ (polar) curves have much lower density compared to the high-$\theta$ (equatorial) curves around the \ac{ns} (vertical line), which is because of the bipolar cavity blown out by the feedback. The profile interior to the location of the \ac{ns} is roughly spherical with low velocities, indicating it is roughly in hydrostatic equilibrium. The exterior is extremely asymmetrical, showing $\lesssim5$ orders of magnitude variations in density depending on the direction. A similar trend can be seen at apastron (Figure~\ref{fig:1d_profile_apastron}), with now a larger hydrostatic interior region and strongly perturbed exterior. Generally, the equatorial profiles seem to have higher densities compared to the polar profiles, showing that the outflow is equatorially concentrated. 

The opacity structure is also quite complex but broadly speaking it is almost like a step function, sharply transitioning from the electron scattering opacity to the floor opacity at the recombination front ($\sim10^4~\rsun$). The radius of the recombination front, which corresponds to the photosphere, varies by a factor $\sim5$ depending on the direction. We note that the opacity jumps at the photosphere are only covered with 3--4 grid points so it is not well resolved. Together with the crude approximation for the opacity that ignores effects of lines, non-equilibrium ionization states, multi-dimensional effects, etc, the temperature and photospheric radius evolution in our models should be approached with caution.

As a visual guide, we overplot the density profile of an example type Ic \ac{sn} ejecta (red curve). See Appendix~\ref{app:1d_explosion} for details of how this was computed. We can see that the bulk of the \ac{sn} ejecta lie at $r\gtrsim2\times10^4~\rsun$ and the density of the inner parts of the ejecta are negligible compared to that of the inflated envelope at radii below that. This indicates that the interaction between the inflated envelope and the inner parts of the \ac{sn} ejecta are weak and will not influence the dynamics in our simulations at least up to the photosphere. Both the inner parts of the \ac{sn} ejecta and the outer parts of the inflated envelope could be compressed into a thin shell at around $\sim10^4~\rsun$, as they interact. However, the mass in this thin shell is expected to be negligible compared to the rest of the \ac{sn} ejecta. As we discuss in Section~\ref{sec:delay_discussion}, the \ac{sn} ejecta may still be optically thick at these times but could become optically thin later on ($\gtrsim50~\mathrm{d}$).

\subsection{Undulation features}

\begin{figure}
 \centering
 \includegraphics[width=\linewidth]{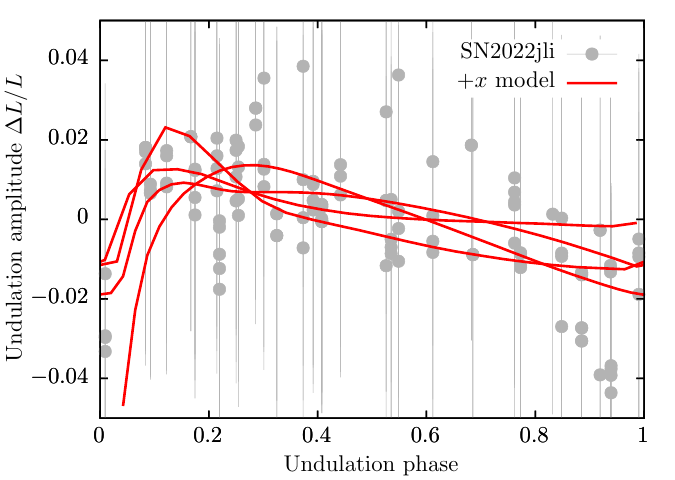}
 \caption{Undulation light curve folded over undulation phase for the M05e07b model. Grey points are the observed data of SN2022jli \citep[]{che24}. Red solid curves are the folded light curves of the $+x$ model in Figure~\ref{fig:lightcurve_e07}. Both are shifted so that the minima are located at the phase origin. \label{fig:undulation_e07}}
\end{figure}

We now compare the shape of the undulations between our model and observations. We focus on the light curve for the $+x$ viewing direction in Figure~\ref{fig:lightcurve_e07} (red curve). To extract the undulations, we first fit a sixth-order polynomial curve to the light curve between 50 and 100~d to obtain a baseline. We plot the difference between the light curve and the baseline against the undulation phase in Figure~\ref{fig:undulation_e07}. The observed undulation light curve is computed in the same way and overplotted. Overall, the model is in good agreement with the observed shape. There is a sharp rise followed by a longer decline. Each cycle has slightly different shapes both in the model and observations, partly due to the ambiguity in defining the baseline, but also due to the stochastic nature of the undulations. Our M05e08b\_r model has similar undulation shapes, although the rise is not as sharp as the lower eccentricity model.

While our edge-on light curve models closely agree with the small amplitude undulations in SN2022jli, we also predict large amplitude oscillations if a similar event was viewed from higher-latitude viewing directions and slightly lower eccentricity. Recently, another rare kind of stripped-envelope \ac{sn} was discovered which re-emerged as a type IIn \ac{sn}, $\sim3$~yr after the first event \citep[SN2022mop;][]{bre25}. In both the tail of the first event and the rise towards the second event, several periodic outbursts ($\sim1$~mag) were detected with $\sim27~\mathrm{d}$ intervals. The shape and amplitude of the outbursts loosely resemble the saw-tooth like morphology that we find in our hydrodynamic models observed from the pole (purple curves in Fig~\ref{fig:lightcurve_e07}). It may be that the first event of SN2022mop was an event similar to SN2022jli, but with different orbital periods and viewed from different angles or had a lower eccentricity. \citet{bre25} report that SN2009ip also has periodic outbursts with amplitudes and intervals similar to SN2022mop.

\section{Discussion}\label{sec:discussion}

\subsection{Super-Eddington accretion}

In our simulations, we do not place a limit on the accretion rate onto the \ac{ns}. Based on the luminosity of the second part of the light curve, the \ac{ns} needs to create a feedback power of about $\sim10^4~L_\mathrm{Edd}$, where $L_\mathrm{Edd}$ is the Eddington luminosity. This exceeds the luminosity of the brightest ultra-luminous X-ray pulsar \citep[NGC 5907 ULX-1;][]{isr17} by an order of magnitude. It has been suggested that such high apparent luminosities can be achieved if the radiation is sufficiently geometrically beamed \citep[]{kin24}. However, in this case the actual feedback power is much less than the isotropic equivalent luminosity of the observed light and thus cannot supply the required energy for our model. Recent general relativistic magneto-hydrodynamics simulations of super-Eddington accretion onto \acp{ns} demonstrate that it is possible at relatively low magnetic field strengths $B\lesssim10^{11}$~G \citep[]{ino23,kay25} to successfully produce radiative and kinetic outflow luminosities reaching $\sim500$--$1000~L_\mathrm{Edd}$, close to what is required for NGC 5907 ULX-1. Such high accretion rates are sustained due to the break in spherical symmetry as the accretion flow is channeled along magnetic field lines towards the magnetic poles. Radiative losses from the sides of the accretion columns cool the matter enough to allow accretion at highly super-Eddington rates.

At such high accretion rates, neutrino emission may start to dominate the cooling in the inner parts of the flow, enabling hypercritical accretion \citep[e.g.][]{che93,che96,fry96}. The accretion rate range for neutrino-dominated hypercritical steady accretion onto \acp{ns} is thought to be around $10^{-4}~\msun~\mathrm{yr}^{-1}\lesssim\dot{M}\lesssim0.1~\msun~\mathrm{yr}^{-1}$, consistent with the accretion rate required to power the luminosity of SN2022jli. In this scenario, the accretion flow is capable of forming a neutrino-cooled disk, which could power bipolar outflows, e.g. via the Blandford-Payne mechanism \citep{bla82}, necessary to avoid excessive negative feedback. However, in this regime it is also expected that a hydrostatic atmosphere is formed \citep{che93,com25}, significantly lowering the accretion feedback luminosity. We note that these neutrino-cooled models tend to ignore the \ac{ns} magnetic field and thus the full picture of neutrino-cooled, magnetically channeled accretion is not understood.

\subsection{$\sim$50~d delay to the second peak}
\label{sec:delay_discussion}

In our synthetic light curves, the accretion feedback occurs straight after the \ac{sn} as the timescale of the companion inflation is on the dynamical timescale. As a result, the accretion feedback kicks in almost immediately, creating noticeable features in the early light curve. Previous studies have shown that the first part of the light curve is consistent with being powered by $^{56}\mathrm{Ni}$ decay and can be modelled as a typical type Ic \ac{sn}. In our hydrodynamic simulations, we have not included the contribution of the \ac{sn} ejecta, which will greatly contribute to the light at early times but also could be optically thick enough to hide the radiation from the accretion feedback. The $\sim50~\mathrm{d}$ delay to the second peak could therefore correspond to the time it takes for the \ac{sn} ejecta to become optically thin and expose the interior binary.

The optical depth of \ac{sn} ejecta can be roughly estimated as
\begin{equation}
 \tau_\mathrm{SN}=\int_0^\infty\kappa\rho_\mathrm{ej}dr=f\bar{\kappa}\frac{M_\mathrm{ej}}{R_\mathrm{ej}^3}R_\mathrm{ej},
\end{equation}
where $\kappa$ is the ejecta opacity and $M_\mathrm{ej}, R_\mathrm{ej}, v_\mathrm{ej}$ are the ejecta mass, radius and velocity, respectively. The factor $f$ is a parameter that depends on the density distribution along the line of sight. Given that we see no signatures of circumstellar matter interaction in SN2022jli, we can safely assume that the ejecta follow a free expanison ($R_\mathrm{ej}=v_\mathrm{ej}t$). The ejecta become optically thin when $\tau_\mathrm{SN}\lesssim1$, at around
\begin{align}
 t_\mathrm{nebular}\gtrsim50~&\mathrm{d}\left(\frac{f}{0.1}\right)^\frac12\left(\frac{\bar{\kappa}}{0.1~\mathrm{g~cm^{-2}}}\right)^\frac12\nonumber\\
&\times\left(\frac{M_\mathrm{ej}}{\msun}\right)^\frac12\left(\frac{v_\mathrm{ej}}{10^4~\mathrm{km~s^{-1}}}\right)^{-1},\label{eq:optical_emergence_time}
\end{align}
which is consistent with the $\sim50~\mathrm{d}$ delay at the order of magnitude level. Indeed, typical stripped-envelope \acp{sn} reach the nebular phase within $\sim100~\mathrm{d}$, roughly consistent with this estimate. For the case of SN2022jli, nebular features start to emerge in the spectrum at around the second peak of the light curve \citep{car24}. This may indicate that the \ac{sn} ejecta is sufficiently optically thin at this point but the spectrum does not become completely nebular due to the emission from the underlying binary.

In our realistic mass accretion model (M05e08b\_r), the light curve shows a slow rise between $30$--$70~\mathrm{d}$, which closely follows the shape of the rise to the second peak in SN2022jli. This slow rise is not seen in our optimistic mass accretion models with lower orbital eccentricity. It could be that the $\sim50~\mathrm{d}$ rise is not due to blanketing by the \ac{sn} ejecta, but a sign that the embedded binary orbit has a high eccentricity ($e\gtrsim0.8$).

\subsection{Rapid drop at $\sim$250~d}

There are several possible explanations for the rapid drop at $\sim250~\mathrm{d}$. First, it could be due to rapid formation of dust, which enshrouds the emitting region. Early dust formation has been detected in some known type Ic \acp{sn} \citep{rho21,rav23}, and evidence of dust formation has been reported for SN2022jli around the time of the drop \citep[]{car24}. Given that the \ac{sn} ejecta are very C-rich (it is a type Ic \ac{sn}), it is a good site for dust formation. As soon as the ejecta adiabatically cool below a threshold temperature, CO molecules can form, which further enhances the cooling through molecular lines. It then quickly reaches the dust condensation temperature to form dust and enshroud the inner binary. If dust was the reason for the drop, we may expect that there is continuous energy injection from the embedded binary and the radiation is reprocessed into infrared bands. Although in SN2022jli the infrared flux does increase after the optical drop ($\sim400~\mathrm{d}$), it does not make up for the decrease in optical flux, indicating a drop in the energy injection is also required.

Another possibility within our scenario is that the envelope inflation ceases at some point, preventing any further mass transfer to the neutron star. The strong accretion onto the \ac{ns} quickly drains the material in the inflated envelope. At some point the \ac{sn}-heated envelope material will be accreted or ejected from the system, bringing an end to the accretion process. Unfortunately due to limitations in computational resources, we were only able to run the simulations up to $\lesssim100~\mathrm{d}$. It will be interesting in the future to extend the simulation to see if such a drop naturally occurs.

We can also consider the so-called propeller effect. The new-born neutron star is likely rapidly spinning and possesses a magnetic field. As the accretion rate onto the neutron star declines over time, there will come a point where the Alfv\'en radius exceeds the corotation radius, beyond which the magnetic field will centrifugally blow away the accreting material. The critical luminosity below which the propeller mechanism kicks in can be estimated by \citep[e.g.][]{cam02}
\begin{align}
 L_\mathrm{lim}\sim3.6\times10^{41}~&\mathrm{erg~s^{-1}}\xi^\frac72\left(\frac{B}{10^{12}~\mathrm{G}}\right)^2\left(\frac{P_\mathrm{spin}}{20~\mathrm{ms}}\right)^{-\frac73}\nonumber\\
&\times\left(\frac{M_\mathrm{NS}}{1.4~\msun}\right)^{-\frac23}\left(\frac{R_\mathrm{NS}}{10~\mathrm{km}}\right)^5,
\end{align}
where $B$ is the strength of the magnetic field, $P_\mathrm{spin}$ the spin period and $\xi$ is an order unity parameter that accounts for geometrical effects. In SN2022jli, the drop at $\sim250~\mathrm{d}$ occurred when the luminosity declined below $\lesssim3\times10^{41}~\mathrm{erg}$, which corresponds to the limiting luminosity of a \ac{ns} rotating at $P_\mathrm{spin}\sim20$~ms for a typical magnetic field strength $B\sim10^{12}$~G. This spin period is within the range of birth spins of NSs predicted from core-collapse simulations \citep[e.g.][]{jan22,bur24}. As soon as the accretion luminosity declines below the critical luminosity ($L_\mathrm{acc}<L_\mathrm{lim}$), the accretion will be magnetically inhibited, leading to a rapid drop in the accretion rate. 

A final possibility is that the infall rate onto the \ac{ns} falls below the threshold for neutrino-dominated accretion. It is unclear what should happen in the transition from a neutrino-dominated to radiative accretion flow, but it is possible that the nature of the accretion changes significantly. For example, as the neutrino cooling weakens, the accretion disk becomes geometrically thicker, making the jet collimation mechanism less efficient. As we see in our simulations, more spherical feedback causes stronger negative feedback, severely quenching the accretion rate.

It is possible that multiple mechanisms are responsible for the drop, as the above models are not mutually exclusive. For example, when the accretion luminosity drops, the temperature around the photosphere will also decline. This could trigger dust formation, further accelerating the drop in optical flux. As we see in later sections, the dust formation is likely a secondary effect and not the sole cause for the drop.

\subsection{Gamma-ray detection}

The origin of the GeV $\gamma$-rays are still unknown. \citet{car24} propose that the $\gamma$-rays could be generated from a magnetar wind nebula. In their model, the nascent \ac{ns} possesses a strong magnetic field ($B\sim8.5\times10^{14}~\mathrm{G}$) as well as rapid spin ($P_\mathrm{spin}\sim50~\mathrm{ms}$), and both the optical light curve and $\gamma$-rays are powered by the spin-down energy.

\begin{figure}
 \centering
 \includegraphics[width=\linewidth]{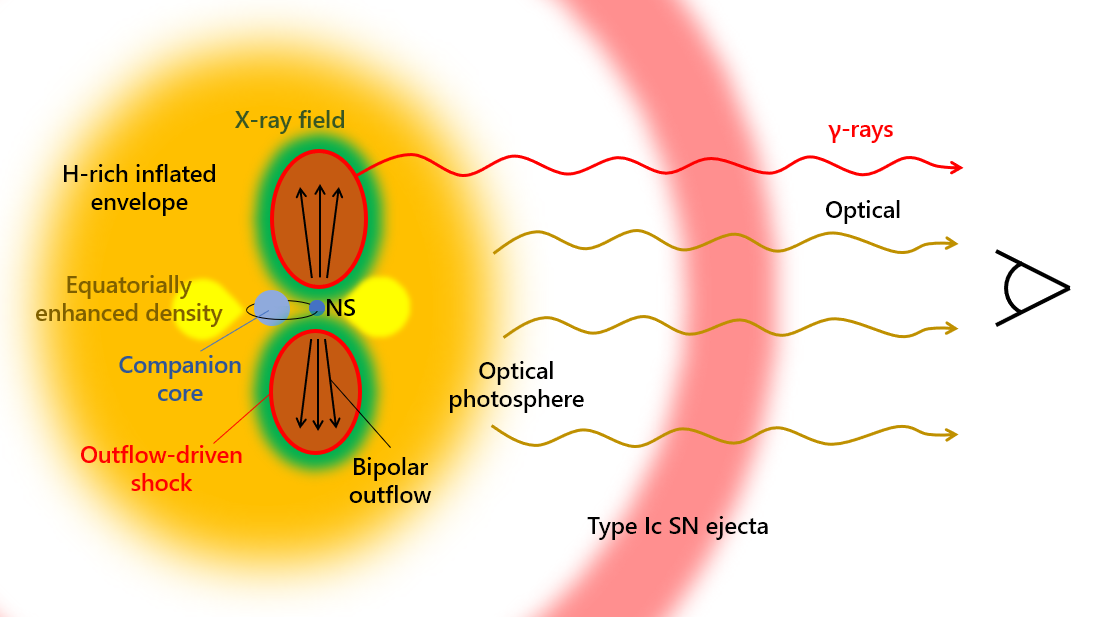}
 \caption{Schematic diagram of the emission from SN2022jli at $\sim200~\mathrm{d}$.\label{fig:schematic}}
\end{figure}

In our accretion-powered scenario, the $\gamma$-rays could be generated  by a relativistic jet launched from the \ac{ns} (Figure~\ref{fig:schematic}).
High-energy ($\sim$100~MeV--TeV) $\gamma$-rays have been observed in the past in stellar mass binaries like LS~I$+61^\circ303$ \citep[]{abd09,alb09}. The compact object in this system is most likely a \ac{ns} \citep[]{wen22}, indicating that \acp{ns} are capable of creating GeV $\gamma$-ray emission. While magnetar-like behaviour has been reported \citep[]{tor12,wen22}, bipolar jets have also been detected in radio \citep[]{mas04,mas12}, complicating the identification of the $\gamma$-ray source in LS~I$+61^\circ303$.

The observed $\gamma$-ray luminosity in SN2022jli is estimated to be $\sim3.1\times10^{41}~\mathrm{erg~s}^{-1}$ in the $1$--$3~\mathrm{GeV}$ band, which is a factor $\sim2$--$3$ lower than the luminosity in the optical. In our light curve models, the optical luminosity is roughly a factor $\sim3$--$6$ lower than the injected luminosity at the accretion peaks (Fig.~\ref{fig:lightcurve_e08}). To have a $\gamma$-ray luminosity $\sim2$--$3$ times lower than this, roughly $\sim5$--$20~\%$ of the accretion luminosity must be converted into $\gamma$-rays.

Both mechanisms (magnetar or relativistic jets) generate $\gamma$-rays at the shock between the \ac{ns} outflow (magnetar wind/accretion-powered jet) and the surrounding material, which in this case is the inflated envelope of the companion. The optical depth of the envelope should decline under $\tau\lesssim1$ for the $\gamma$-rays to emerge. The $\gamma$-rays were only detected at $\sim200~\mathrm{d}$, meaning that it must have been obscured before that. For GeV photons, the main source of opacity is the Bethe-Heitler pair production process \citep{vur21,lu25}. The effective opacity is given by \citep[]{zdz89,vur21}
\begin{equation}
 \kappa_{\gamma p}=\frac{21}{2\pi}\alpha_\mathrm{fs}\left(\ln{\left[\frac{2\epsilon_\gamma}{m_ec^2}\right]}-\frac{109}{42}\right)\kappa_\mathrm{T},
\end{equation}
where $\alpha_\mathrm{fs}\sim1/137$ is the fine structure constant, $\epsilon_\gamma$ is the photon energy, $m_e$ is the electron mass, $c$ the speed of light and $\kappa_\mathrm{T}$ is the Thomson scattering opacity. For GeV photons, this becomes $\kappa_{\gamma p}\sim0.03~\mathrm{g~cm^{-2}}$. Following Eq.~(\ref{eq:optical_emergence_time}), we can estimate the $\gamma$-ray emergence time as 
\begin{align}
 t_{\gamma,\mathrm{ej}}\gtrsim30~&\mathrm{d}\left(\frac{f}{0.1}\right)^\frac12\left(\frac{M_\mathrm{ej}}{\msun}\right)^\frac12\left(\frac{v_\mathrm{ej}}{10^4~\mathrm{km~s^{-1}}}\right)^{-1}.\label{eq:gamma_emergence_time}
\end{align}
The ejecta would thus become transparent to GeV photons within months, although it may be compatible with $\sim200~\mathrm{d}$ if the ejecta mass was larger and the associated ejecta velocity was slower.

The shocks responsible for generating $\gamma$-rays could also create X-ray emission via e.g. synchrotron radiation. These photons could act as a source of opacity through photon-photon absorption, in which two photons interact to create electron-positron pairs ($\gamma+\gamma\rightarrow e^-+e^-$). To absorb GeV $\gamma$-ray photons, the target soft photons should have an energy $\epsilon_\mathrm{soft}\geq(2m_ec^2)^2/\epsilon_\gamma\sim1~\mathrm{keV}$. Although such keV photons will eventually be re-processed into optical wavelengths via Thomson scattering, they could still interact with the $\gamma$-rays before being diminished. The optical depth for photon-photon absorption can be estimated as \citep[]{bar11}
\begin{equation}
 \tau_{\gamma\gamma}\sim\frac{\sigma_T}{5} \frac{L_\mathrm{soft}}{4\pi R_\mathrm{sh}c\epsilon_\mathrm{soft}},
\label{eq:gg_optical_depth}
\end{equation}
where $L_\mathrm{soft}$ is the X-ray luminosity, $R_\mathrm{sh}$ is the radius of the $\gamma$-ray emitting shock front and $\epsilon_\mathrm{soft}=1~\mathrm{keV}$ is the X-ray energy.
In this scenario, GeV emission will appear when $R_\mathrm{sh}$ exceeds the ``absorption radius''
\begin{equation}
R_\mathrm{abs}\sim\frac{\sigma_T}{5} \frac{L_\mathrm{soft}}{4\pi c \epsilon_\mathrm{soft}} \approx 3000~\rsun\left(\frac{L_\mathrm{soft}}{10^{42}~\mathrm{erg~s}^{-1}}\right),
\end{equation}
below which the GeV photons will be absorbed by the keV photons.
In our simulations, the polar shock front lies roughly around $R_\mathrm{sh}\sim2000~\rsun$ at $\sim50~\mathrm{d}$ (see the outer edge of the polar $v\sim10^{10}~\mathrm{cm~s}^{-1}$ component in Figures~\ref{fig:1d_profile_periastron} \& \ref{fig:1d_profile_apastron}) and advances linearly in time. If we assume that the X-ray luminosity at the shock is equal to the observed optical luminosity, the absorption radius declines over time. The crossover of the increasing $R_\mathrm{sh}$ and decreasing $R_\mathrm{abs}$ occurs in a few months, not too far from the $\sim200~\mathrm{d}$ delay in SN2022jli. See Figure~\ref{fig:schematic} for a schematic diagram of the $\gamma$-ray emission in our model.

There are no reported detections of $\gamma$-rays after the rapid optical drop. If accretion is ongoing, we would expect the $\gamma$-rays to still be observable after the drop because dust is mostly transparent to $\gamma$-rays. The lack of ongoing $\gamma$-ray emission indicates that dust obscuration cannot be the sole reason for the optical drop but the accretion has likely terminated or significantly weakened around the drop.

\subsection{H$\alpha$ emission}
\label{sec:Halpha}

H$\alpha$ emission has been detected in the late-time spectrum, showing cyclic shifting patterns. \citet{che24} propose that the shifting could be consistent with the orbital motion of the surviving companion star on a highly eccentric orbit. Based on their comparison to the radial velocity curve, the shifting is roughly consistent with an orbital eccentricity of $e\sim0.7$--$0.96$. Their suggested eccentricity range is in agreement with our hydrodynamic models, which show that a relatively high eccentricity of $e\gtrsim0.8$ is required to reach sufficiently high accretion rates.

It is still questionable whether the cyclic pattern they identify are due to the orbital motion of the companion. The velocity profile is rather complex with multiple peaks and the wings spread out to $|v_{\mathrm{H}\alpha}|\sim4000~\mathrm{km~s}^{-1}$. In fact, there are no obvious peaks in the orbital velocity range inferred from the flux-weighted centroid \citep[$|v_{\mathrm{H}\alpha}|<600~\mathrm{km~s}^{-1}$;][]{che24}. Instead, there appears to be a steady broad component with a maximum at around $v_{\mathrm{H}\alpha}\sim+1000$--$2000~\mathrm{km~s}^{-1}$ and another component at $v_{\mathrm{H}\alpha}\sim-1000~\mathrm{km~s}^{-1}$ that strengthens around the undulation peaks \citep[]{car24}.

In our models, the photosphere has two components: a steadily increasing baseline and distinct spikes at periastron (Figure~\ref{fig:Teff_Rphot_e07}). Very roughly, the photospheric velocity is $\sim1000~\mathrm{km~s}^{-1}$. Around periastron, bipolar outflows are ejected at $\sim10^4~\mathrm{km~s}^{-1}$, quickly penetrating the photosphere. It may be possible that the bipolar ejecta are creating the $-1000~\mathrm{km~s}^{-1}$ H$\alpha$ emission while the $+1000~\mathrm{km~s}^{-1}$ is created by the baseline photosphere. More detailed spectroscopic modelling is required to firmly constrain the nature of the H$\alpha$ emission.

\subsection{Orbital eccentricity}
\label{sec:eccentricity}

\begin{figure}
 \centering
 \includegraphics[width=\linewidth]{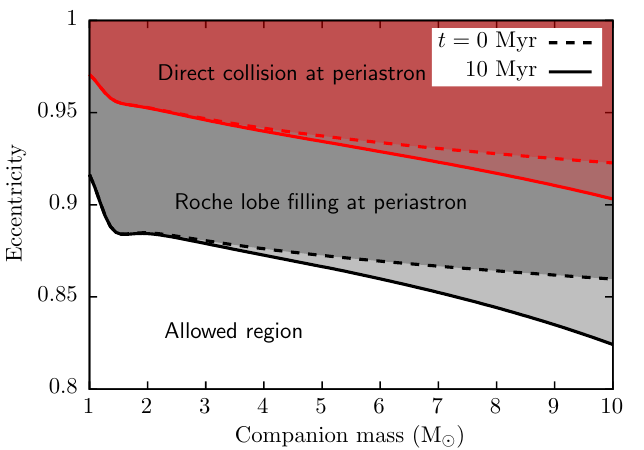}
 \caption{Allowed eccentricities as a function of companion mass. The neutron star mass is fixed to $M_\mathrm{NS}=1.4~\msun$ and the orbital period to $P_\mathrm{orb}=12.5~\mathrm{d}$. Black curves show the minimum eccentricity above which the companion star will overflow its effective Roche lobe at periastron whereas the red curves show the minimum eccentricity above which the \ac{ns} will plunge into the companion at periastron. We use the \citet{hur00} formulae to estimate the stellar radius $R_2$ at zero-age main sequence (dashed) and at $t=10~\mathrm{Myr}$ (solid).\label{fig:emax}}
\end{figure}

Within our scenario, we can further narrow down the possible range of eccentricity based on orbital stability. The undulation period in SN2022jli is preserved for at least $\gtrsim200~\mathrm{d}$, indicating that there is no orbital decay through drag or mass transfer processes. Figure~\ref{fig:emax} displays the maximum eccentricity to avoid orbital decay, computed as $e_\mathrm{max}=1-R_2/(fa)$. For the red curves we set $f=1$ ($a_\mathrm{per}=a(1-e_\mathrm{max})=R_2$) and for the black curves we use the Roche lobe radius formula \citep[]{egg83}
\begin{equation}
 f=\frac{0.49q^{2/3}}{0.6q^{2/3}+\log(1+q^{1/3})},
\end{equation}
where $q=M_2/M_\mathrm{NS}$ is the mass ratio. Any eccentricity in the grey shaded region would cause Roche lobe overflow at periastron, likely leading to noticeable changes in the orbital period. The red region is a harder limit where the \ac{ns} will plunge into/through the companion star core, which should drive rapid orbital decay \citep[]{RH22b}. Given that there is no evidence of orbital decay all the way to the drop, the system eccentricity should have been below the grey shaded region ($e\lesssim0.88$).

\subsection{Future evolution}

The inflation of \ac{sn}-heated companion stars only last for several years to decades \citep[]{RH15,RH18,oga21,RH23}. Once the star regains thermal equilibrium, it should return to a regular main-sequence star which does not overflow its Roche lobe. Our hydrodynamic simulations unfortunately cannot place strong constraints on the possible companion mass of SN2022jli (Figure~\ref{fig:mdot_mass}). Here, we speculate on the future of SN2022jli depending on what the companion star mass is.

If the companion star is an OB-type star ($M_2\gtrsim8~\msun$), the system may immediately transition into an eccentric wind-fed high-mass X-ray binary similar to GX 301-2 \citep[]{sat86}. Given that there was a likely mass transfer phase preceding the \ac{sn} that stripped the hydrogen and helium layers, the companion star could have accreted a substantial amount of mass and angular momentum. Therefore, the companion may eventually evolve into a rapidly-rotating Oe/Be star, turning the system into a Be X-ray binary like LS~I$+61^\circ303$. This system shares many properties with the SN2022jli system, such as a similar orbital period \citep[$P_\mathrm{orb}\sim26.5~\mathrm{d}$;][]{tay82}, eccentricity \citep[$e\sim0.54$;][]{ara09} and high-energy $\gamma$-ray emission \citep[]{abd09,alb09}. In this case, SN2022jli was marking the birth of a high-mass X-ray binary, with an enhanced accretion phase straight after the \ac{sn}.

If the companion is a lower mass star ($M_2\sim2$--$5~\msun$), the system will soon become dormant, as there will be neither Roche lobe overflow nor sufficiently strong winds to maintain accretion. In this detached phase, it could be observed as an astrometric binary with a dark companion \citep[e.g. Gaia NS1;][]{elb24a,elb24b}. It will only re-activate once the companion evolves to emit strong winds and become a symbiotic X-ray binary or overflow its Roche lobe and transition into a low/intermediate-mass X-ray binary phase. A system like Circinus X-1 could be a possible end product, which again shares many similarities with SN2022jli such as the orbital period and eccentricity \citep[$P_\mathrm{orb}\sim16.5~\mathrm{d}, e\sim0.45$;][]{jon07}. The current X-ray light curve bears a striking similarity to the SN2022jli optical light curve, with saw-tooth-like modulations \citep[]{tom23}. It is also known to launch relativistic jets \citep[]{fen04}. It is possible that SN2022jli was a Cir X-1-like system embedded in an inflated envelope, which eventually transitioned out of the accretion phase but will return in the future.

\section{Summary and conclusion}\label{sec:summary}

We present 3D hydrodynamic simulations of \ac{sn}-induced binary-interaction-powered \acp{sn} and predict their observational features. We find that our scenario can robustly produce light curves with periodic undulations and a steady decline in the long term. The luminosity ($\sim$ mass accretion rate) depends strongly on various factors, such as the form of accretion feedback and orbital eccentricity.

Based on our parameter study, a high-eccentricity ($e\gtrsim0.8$) orbit with geometrically confined feedback is required to reach mass accretion rates compatible with the luminosity of SN2022jli. The viewing angle dependence is insignificant at these high eccentricities, but become increasingly important at $e\lesssim0.7$. Higher viewing angle models with lower eccentricity are reminiscent of more strongly outbursting \acp{sn} like SN2022mop or SN2009ip. On the other hand, the companion mass does not seem to influence the accretion rate significantly.

Binary-interaction-powered \acp{sn} provide strong evidence that binary interactions like mass transfer or common envelope phases are responsible for stripping the hydrogen-rich envelope and explode as stripped-envelope supernovae \citep[]{pod92}.
By comparing our models to the observed properties of the light curve, we can further constrain the immediate post-\ac{sn} binary properties. Similar to how post-\ac{sn} companion detections can help constrain binary interaction physics \citep[]{RH23}, the post-\ac{sn} binary properties can also indirectly constrain binary interaction physics. For example, in our model for SN2022jli, the post-\ac{sn} periastron distance must be $a_\mathrm{per}\lesssim19~\rsun$. This directly relates to the pre-\ac{sn} orbital separation, which in turn relates to the post-binary interaction orbital separation. If the system had undergone a common-envelope phase, our constraint can provide useful insight into the long-sought post-common envelope separations in massive binary systems.

Since the first discovery of periodic undulations in a \ac{sn} (SN2022jli), a growing number of possible periodic outbursts have been reported in the past few years, e.g. SN2022mop, SN2009ip \citep[]{bre25}, SN2015ap \citep[]{rag25}. These all may be marking the birth of compact object binaries including X-ray binaries, detached \ac{ns} binaries, etc. Many more such \acp{sn} are expected to be discovered in the coming years with next-generation transient survey telescopes including the Vera Rubin Observatory and the Wide Field Survey Telescope. In particular, the high cadence of the Legacy Survey of Space and Time (LSST) by the Rubin Observatory (especially in the deep drilling fields) could be extremely powerful for identifying periodic signatures in \ac{sn} light curves. Our models should serve as a first guide into what to expect in the future, and how to infer the underlying binary properties from the observed light curves.

%% Also note that the akcnowlodgment environment does not support long amounts of text. If you have a lot of people and institutions to acknowledge, do not use this command. Instead, create a new \section{Acknowledgments}.
\begin{acknowledgments}
The authors thank the anonymous referee for constructive comments that significantly improved the quality of the work. RH thanks Akihiro Inoue for useful discussions on super-Eddington accretion.
This work was supported by software support resources awarded under the Astronomy Data and Computing Services (ADACS) Merit Allocation Program. ADACS is funded from the Astronomy National Collaborative Research Infrastructure Strategy (NCRIS) allocation provided by the Australian Government and managed by Astronomy Australia Limited (AAL). 
This work was partially performed on the OzSTAR national facility at Swinburne University of Technology. The OzSTAR program receives funding in part from the Astronomy NCRIS allocation provided by the Australian Government. 
Part of the simulations were performed on the supercomputer Gadi with the assistance of resources from the National Computational Infrastructure (NCI Australia), an NCRIS enabled capability supported by the Australian Government. 
Part of the simulations were also performed on the HOKUSAI computing facility in RIKEN.
This work was initiated in part at Aspen Center for Physics, which is supported by National Science Foundation grant PHY-2210452. 
S.N. is supported by JSPS Grant-in-Aid Scientific Research (KAKENHI) (A), Grant Number JP25H00675 and (B), Grant Number JP23K25874, and by JST ASPIRE Program ``RIKEN-Berkeley mathematical quantum science initiative''.
\end{acknowledgments}


\vspace{5mm}
%\facilities{HST(STIS), Swift(XRT and UVOT), AAVSO, CTIO:1.3m,CTIO:1.5m,CXO}


\software{HORMONE \citep{RH16,RH22b},
          MESA \citep{MESA1,MESA2,MESA3,MESA4,MESA5},
          MESA SDK \citep{mesasdk},
          gnuplot \citep{Gnuplot}
          }



\appendix

\section{Approximate opacity formula}\label{app:opacity}

For the opacity in our light curve calculations, we use an approximate analytical formula similar to that of \citet{met17}. We combine various approximate formulae as
\begin{equation}
 \kappa=\max\left[\kappa_\mathrm{floor}, \kappa_\mathrm{mol}+\left(\kappa_{\mathrm{H}^-}^{-1}+\left(\kappa_\mathrm{es}+\kappa_\mathrm{K}\right)^{-1}\right)^{-1}\right],\label{eq:opacity}
\end{equation}
where the electron scattering opacity is
\begin{align}
 \kappa_\mathrm{es}=&~0.2(1+X)\left(1+2.7\times10^{11}\frac{\rho/\mathrm{(g~cm^{-3})}}{(T/\mathrm{K})^2}\right)^{-1}\nonumber\\
&\times\left(1+\left(\frac{T}{4.5\times10^8~\mathrm{K}}\right)^{0.86}\right)^{-1}~\mathrm{cm^2~g^{-1}},
\end{align}
Kramers opacity (bound-free/free-free absorption) is expressed as
\begin{equation}
 \kappa_\mathrm{K}\approx4\times10^{25}Z(1+X)\left(\frac{\rho}{\mathrm{g~cm^{-3}}}\right)\left(\frac{T}{\mathrm{K}}\right)^{-7/2}~\mathrm{cm^2~g}^{-1},
\end{equation}
 we approximate the opacity of negative hydrogen as
\begin{equation}
 \kappa_{\mathrm{H}^-}\approx1.1\times10^{-40}Z^{0.5}\left(\frac{\rho}{\mathrm{g~cm^{-3}}}\right)^{0.2}\left(\frac{T}{\mathrm{K}}\right)^{11}~\mathrm{cm^2~g^{-1}},
\end{equation}
and $\kappa_\mathrm{mol}\approx0.1Z~\mathrm{cm^2~g^{-1}}$ is the molecular opacity.
The form of $\kappa_\mathrm{H^-}$ was chosen by a fit-by-eye approach to the opacity table presented in \citet{far24}. We do not include dust opacities, as dust formation was only detected after $\gtrsim250~\mathrm{d}$ \citep[]{car24}. A floor opacity $\kappa_\mathrm{floor}$ is set to account for various other effects including lines, velocity gradients, non-\ac{lte} effects, etc.

\begin{figure}
 \centering
 \includegraphics[width=\linewidth]{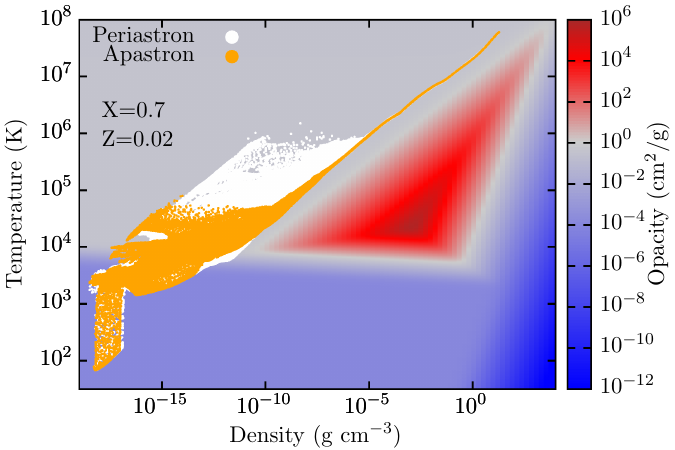}
 \caption{Opacity as a function of density and temperature computed from Eq.~(\ref{eq:opacity}) with $\kappa_\mathrm{floor}=0$. Overplotted are the density and temperature of each cell in our simulations from the two snapshots used in Figures~\ref{fig:1d_profile_periastron} \& \ref{fig:1d_profile_apastron}.\label{fig:opacity_landscape}}
\end{figure}

The resulting opacities are displayed in Figure~\ref{fig:opacity_landscape}. We find that for the relevant density--temperature combinations in our simulation, the opacity is dominated by electron scattering.

\section{1D explosion simulation}\label{app:1d_explosion}
To roughly represent the density profile of type Ic \ac{sn} ejecta, we employ a thermal bomb technique to explode a stellar model of a naked CO core. We first evolve a $15~\msun$ star in MESA, mostly adopting the inlists in the \texttt{20M\_pre\_ms\_to\_core\_collapse} test suite. After core He depletion, we reduce the mass down to $3~\msun$, stripping off the H and He-rich layers. This exposes the CO-rich layers to the surface, representing a type Ic \ac{sn} progenitor. The star is further evolved up to core collapse. We then map the MESA stellar profile into a 1D spherical coordinate grid in HORMONE, excising the inner $1.5~\msun$ from the computational domain to represent the proto neutron star. The radial direction is split into 1500 grid points where the grid spacing is increased outwards in a geometrical series and the inner most cell size is $\Delta r_1=10^7~\mathrm{cm}$. At $t=0$, we inject $E_\mathrm{inj}=1.18\times10^{51}~\mathrm{erg}$ in the innermost 10 cells to drive an explosion. Subtracting off the envelope binding energy, this leads to an ejecta kinetic energy of $E_\mathrm{exp}=10^{51}~\mathrm{erg}$. A low-density atmosphere is attached outside the star that has negligible mass compared to the star. A reflective boundary is applied at the inner boundary and an outgoing boundary is applied at the outer boundary.

As soon as the simulation is started, a strong outgoing shock is formed around the energy-added cells. We follow the evolution until the shock breaks out of the stellar surface and reaches a state of homologous expansion. For the red curves in Figures~\ref{fig:1d_profile_periastron} \& \ref{fig:1d_profile_apastron}, we take the final snapshot and appropriately rescale the density profile assuming homologous expansion.

\end{CJK*}
\end{document}